\documentclass[%
 aip,
 amsmath,amssymb,
 reprint,%
 floatfix,
]{revtex4-2}

\usepackage{graphicx}
\usepackage{xcolor}
\colorlet{RED}{red}

\colorlet{splitgreen}{green!50!black}
\colorlet{SPLITGREEN}{splitgreen}

\usepackage[english]{babel}
\usepackage{amsmath}
\usepackage[colorlinks=true, allcolors=blue]{hyperref}

\begin{document}
\title{Multistability in the Duffing--Holmes oscillator: basins of
antiperiodic, periodic, and chaotic attractors and their relative volumes}

\author{Arturo C. Marti}
\email{marti@fisica.edu.uy}
\affiliation{Facultad de Ciencias, Universidad de la Rep\'ublica,
Igu\'a 4225, Montevideo 11400, Uruguay}

\author{E.~D.~Leonel}
\affiliation{Departamento de F\'isica, Universidade Estadual Paulista
             (UNESP), Av.\ 24A 1515, Rio Claro, S\~ao Paulo, Brazil}

\date{\today}

  \begin{abstract}The periodically driven Duffing--Holmes oscillator is strongly multistable:
for the same forcing amplitude and frequency, periodic, chaotic, and
antiperiodic attractors coexist, each with its own basin of attraction.
Antiperiodic orbits are periodic orbits invariant under a shift of half a
driving period combined with a sign reversal of the coordinates, a symmetry
of the equations of motion. Guided by a continuation map of the parameter
plane, which locates the regimes but is blind to their coexistence, we
explore the space of initial conditions along two one-parameter cuts, at
fixed forcing frequency and at fixed forcing amplitude. Stochastic sampling
of the full space yields an unbiased estimate of the relative volume of each
basin as a function of the control parameter, on the same parameter grids on
which single-trajectory bifurcation diagrams are computed by continuation.
Basin maps on two-dimensional sections, complemented by maps of the order of
the attractor reached from each initial condition, resolve the geometry of
the coexisting basins where the competition between attractors is strongest.
The relative volumes reveal windows of global antiperiodic dominance, in
which the antiperiodic orbit attracts essentially every initial condition,
separated by narrow intervals in which two or three qualitatively different
attractors share the space in comparable proportions, as well as extended
ranges of persistent coexistence; at large forcing amplitude, a single
antiperiodic orbit of order one captures the whole space. Compared with the
bifurcation diagrams, the abrupt changes of basin volume are the
ensemble-level counterpart of the multistability-induced discontinuities of
the sweeps, and the gradual shrinking of the antiperiodic basin explains why
the branch followed by a sweep depends on the initial state and on the drive
phase.
\end{abstract}

\maketitle

\begin{quotation}
A nonlinear system driven far from equilibrium may have several
long-term behaviors available for the same values of its parameters;
which one is realized depends on where the system starts. This
multistability turns the question ``what does the system do?'' into
``how likely is each of its possible fates?'', a question about the
relative sizes of the basins of attraction. Here we answer it for one of
the simplest chaotic systems, a particle in a symmetric double-well
potential shaken periodically, whose stable responses include, besides
periodic and chaotic oscillations, antiperiodic ones, which reproduce
their own negative after a fixed delay. By sampling the space of initial
conditions at random we measure how the coexisting attractors share it
as the forcing is varied, and find that the antiperiodic response often
captures the whole space, that the competition between attractors is
confined to narrow parameter windows, and that in those windows the
outcome of a slow parameter sweep is decided by the initial state,
including the phase of the drive, rather than by the parameters alone.
\end{quotation}

\section{Introduction}
\label{sec:intro}

Multistability, the coexistence of several attractors for the same
values of the parameters, is a generic feature of nonlinear dissipative
systems, documented in lasers, chemical reactions, neuronal and
ecological models, and mechanical
oscillators~\cite{feudel2008complex,pisarchik2014control}. Its practical
consequences follow from a single fact: the asymptotic state of the
system is no longer determined by the parameters alone but also by the
initial condition, through the basin of attraction in which it lies.
When basin boundaries are fractal, the final state becomes unpredictable
for initial conditions specified with finite
precision~\cite{grebogi1983final,mcdonald1985fractal,aguirre2009fractal}; when a parameter
is varied slowly, the system may jump abruptly from one attractor to
another, with hysteresis on the return path.

A quantitative description of multistability therefore requires, beyond
the identification of the coexisting attractors, a characterization of
their basins: their geometry and the uncertainty of their boundaries,
quantified for instance by the basin
entropy~\cite{daza2016basin,daza2023unpredictability}, and their
relative volumes, which measure the probability that a random initial
condition ends on a given attractor and constitute a natural global
measure of its stability~\cite{menck2013basin}. The latter are estimated
by random sampling of the initial conditions, and their dependence on a
control parameter has become a tool of global stability analysis in its
own right~\cite{gelbrecht2020monte,stender2022bstab,datseris2023framework};
the variety of basin structures found in low-dimensional systems has
recently been reviewed~\cite{wagemakers2025basins}.

The periodically driven Duffing--Holmes oscillator, a damped particle in
a symmetric double-well potential subjected to a harmonic force, is one
of the simplest continuous-time systems in which all these phenomena
occur~\cite{holmes1979nonlinear,ueda1980,ChaosKorsch,strogatz2018nonlinear,argyris2015exploration}.
Fractal basin boundaries were documented in this very system in the
early days of the subject~\cite{moon1985fractal}, the abrupt
reorganizations of its basins as the forcing is varied were traced to
global bifurcations of homoclinic and heteroclinic
type~\cite{ueda1990basin}, its basin boundaries were later shown to
possess the Wada property~\cite{aguirre2002unpredictable}, its parameter
plane contains intricate periodic and chaotic
domains~\cite{bonatto2008chaotic} and fractal boundaries between
single-well and cross-well responses~\cite{hasan2023fractal}, and both
its basin structure~\cite{todd1997experimental} and the abrupt jumps
between coexisting attractors along a parameter sweep have been observed
experimentally, in a mechanical analogue and in analog electronic
implementations~\cite{gargiulo2026order}.

The double-well geometry adds a feature absent from single-well
oscillators: an attractor may be confined to one well or visit both, and
the two wells are exchanged by a symmetry of the equations of motion, so
that single-well attractors necessarily occur in conjugate pairs with
conjugate basins. In a companion paper~\cite{marti2026antiperiodicity},
hereafter referred to as Ref.~I, we studied the organization of the
periodic, chaotic, and antiperiodic regimes of this oscillator in the
plane spanned by the forcing amplitude and frequency. Antiperiodic
orbits, satisfying $x(t+T)=-x(t)$ for some antiperiod $T$, were shown
there to be precisely the periodic orbits invariant under a discrete
symmetry $S$ of the equations of motion, defined in
Sec.~\ref{sec:model} below, which exchanges the wells while shifting
time by half a forcing period; every other periodic orbit breaks this
symmetry and occurs as one member of a conjugate pair.

Invariance under $S$ forces antiperiodic orbits to lock to the drive at
odd multiples of its period and to visit both wells, and restricts the
ways in which they can lose stability to two: a symmetry-breaking
pitchfork, in which the antiperiodic orbit gives way continuously to a
conjugate pair of the same order, or a saddle-node bifurcation with an
antiperiodic saddle, after which the system jumps to an unrelated
attractor. Those results were obtained by following a single attractor
by numerical continuation as one parameter is varied. Multistability
entered that description only indirectly, through the abrupt jumps
between branches, the multistability-induced discontinuities, which such
sweeps display wherever the tracked attractor loses its stability or its
basin, and whose outcome depends on the initial state and on the phase
of the drive.

Here we focus on the multistability itself and on its relation with the
symmetry $S$. The question is how the coexisting antiperiodic, periodic,
and chaotic attractors share the space of initial conditions, and how
the symmetry, which distinguishes the antiperiodic orbits from all the
others, is reflected in their basins. An overview of the parameter
plane, computed by continuation over a region wider than the one mapped
in Ref.~I (Sec.~\ref{sec:plane}), locates the regimes and two
one-parameter cuts, at fixed forcing amplitude and at fixed forcing
frequency, along which the coexistence is then resolved.

Along these cuts we explore the three-dimensional space of initial
conditions, position, velocity, and initial phase of the drive, in two
ways. Stochastic sampling of the full space (Sec.~\ref{sec:fractions})
yields an unbiased estimate of the relative volume of every basin as a
function of the control parameter, on exactly the same parameter grids
on which the bifurcation diagrams of a single trajectory are recomputed
by continuation, so that the two descriptions can be compared value by
value. Basin maps on two-dimensional sections (Sec.~\ref{sec:basins}),
together with maps of the order of the attractor reached from each
initial condition, then resolve the geometry of the coexisting basins at
the values of each cut at which the competition between attractors is
strongest.

The comparison between the relative basin volumes and the
single-trajectory diagrams (Sec.~\ref{sec:discussion}) is the main
outcome of the paper: it identifies which discontinuities of the
continuation sweeps correspond to a genuine competition between
attractors with substantial basins, distinguishes the ensemble-level
signatures of the two ways in which an antiperiodic window can end, and
shows that over extended parameter ranges the antiperiodic orbit is not
merely present but globally dominant. Section~\ref{sec:model}
summarizes the model and the symmetry, Sec.~\ref{sec:conclusions}
collects the conclusions, and the numerical methods, with emphasis on
the sampling procedure, are described in Appendix~\ref{app:methods}.

\section{Model, symmetry, and dynamical regimes}
\label{sec:model}

The Duffing--Holmes oscillator is described by
\begin{equation}
    \ddot{x} + \delta\,\dot{x} + \alpha\,x + \beta\,x^{3}
    = \gamma \cos(\omega t + \phi_{0}),
    \label{eq:duffing}
\end{equation}
where $\delta\geq0$ is the damping coefficient, $\alpha$ and $\beta$
control the linear and cubic restoring forces, $\gamma$ and $\omega$ are
the amplitude and angular frequency of the drive, and $\phi_{0}$ is its
initial phase. Throughout this work we fix $\alpha=-1$, $\beta=1$, and
$\delta=0.3$, so that the conservative part of Eq.~\eqref{eq:duffing}
derives from the symmetric double-well potential
$V(x)=\tfrac{1}{2}\alpha x^{2}+\tfrac{1}{4}\beta x^{4}$, with minima at
$x=\pm1$ separated by the barrier at $x=0$; the line $x=0$ is used
throughout to decide whether a trajectory remains confined to a single
well or visits both. These values are common to all the results
presented below and are not repeated in the captions.

The phase space of Eq.~\eqref{eq:duffing} is three-dimensional, with
coordinates $(x,\dot{x},\theta)$, where $\theta=\omega t+\phi_{0}$ is
taken modulo $2\pi$. An initial condition is thus a point
$(x(0),\dot{x}(0),\phi_{0})$ of this space: the initial phase of the
drive is a coordinate of the initial state, of the same standing as the
position and the velocity, and not a parameter of the system. The basins
of attraction studied below are subsets of this three-dimensional space.

Equation~\eqref{eq:duffing} is invariant under the half-period shift
symmetry
\begin{equation}
  S:(x,\dot{x},t)\mapsto(-x,-\dot{x},\,t+T_d/2), \qquad T_d=\frac{2\pi}{\omega},
  \label{eq:S}
\end{equation}
which exchanges the two wells and shifts time by half a forcing period;
on the space of initial conditions it acts as
$(x(0),\dot{x}(0),\phi_{0})\mapsto(-x(0),-\dot{x}(0),\phi_{0}+\pi)$.
The consequences of this symmetry for the attractors of
Eq.~\eqref{eq:duffing} follow from the general theory of symmetric
periodically forced
systems~\cite{swift1984suppression,chossat1988symmetry,melbourne1993structure}
and were worked out in Ref.~I. Every attractor is either invariant under
$S$ or one member of a conjugate pair $A$, $B=S(A)$ with
$A\cap B=\emptyset$. For a periodic orbit, invariance under $S$ is
antiperiodicity, $x(t+T_{\mathrm{orb}}/2)=-x(t)$, and it requires the
orbital period to be an odd multiple of the driving period,
$T_{\mathrm{orb}}=kT_d$ with $k$ odd; an antiperiodic orbit therefore
visits both wells. Periodic orbits that break the symmetry occur as
conjugate pairs; single-well orbits, which are never $S$-invariant, are
the most common instance. For a chaotic attractor, invariance under $S$
constrains only the attractor as a set: a chaotic attractor visiting
both wells may be $S$-invariant, whereas single-well chaos always occurs
in conjugate pairs.

Two properties of the basins follow directly and are used below. Since
$S$ maps trajectories onto trajectories, the basin of $B=S(A)$ is the
image under $S$ of the basin of $A$, so that conjugate basins have
exactly the same volume within any $S$-invariant region of the space of
initial conditions; and the basin of an $S$-invariant attractor is
itself an $S$-invariant set (for the general relation between the
symmetry groups of an attractor, of its basin, and of its basin boundary
in equivariant systems, see Ref.~\cite{xie2025symmetry}).
Note that a section of the
space of initial conditions at fixed drive phase is not $S$-invariant,
since $S$ maps the section $\phi_{0}=0$ onto the section $\phi_{0}=\pi$,
so that the equality of conjugate basins holds for their volumes in the
full space but not, in general, for their areas on such a section.

\begin{figure*}[!t]
  \centering
  \includegraphics[width=\textwidth]{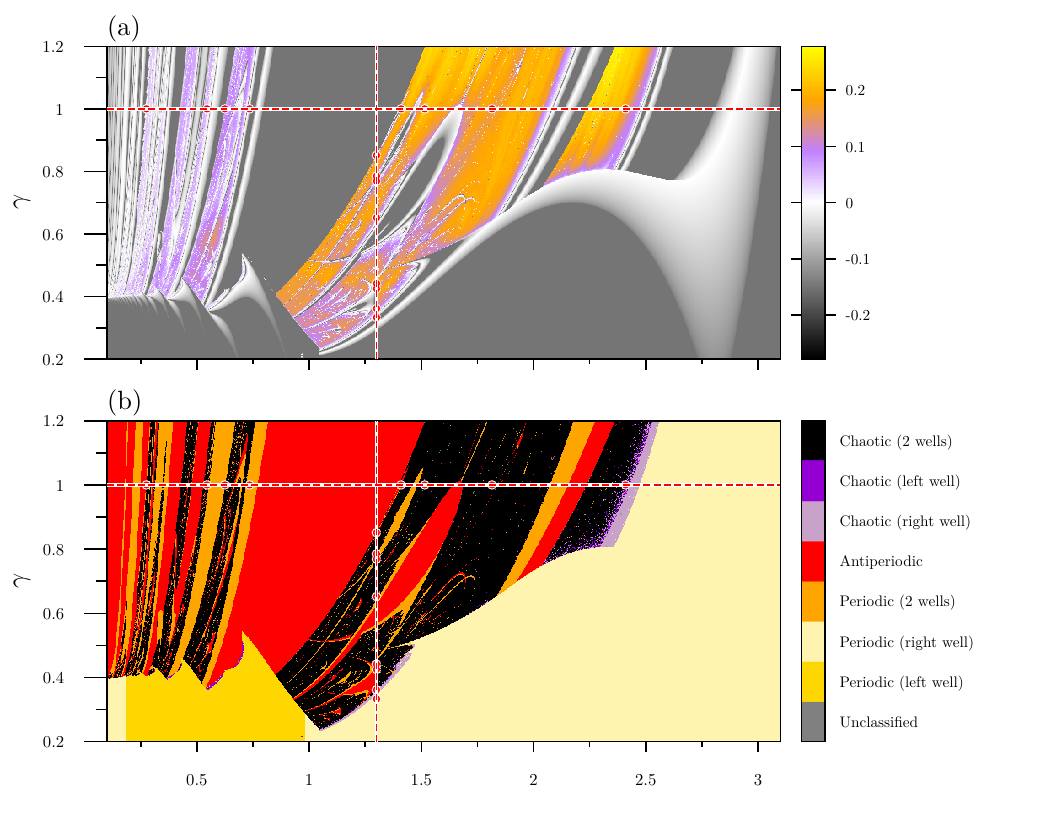}
  \caption{\label{fig:plane_regimes}%
  (a)~Maximum Lyapunov exponent $\lambda_{\max}$ and (b)~dynamical regime
  of the attractor reached by continuation in the $(\omega,\gamma)$ plane,
  $\omega\in[0.1,3.1]$, $\gamma\in[0.2,1.2]$, on an $800\times800$ grid.
  Along each column (fixed $\omega$) the integration is seeded at
  $\gamma=0.2$ with $\bigl(x(0),\dot{x}(0),\phi_{0}\bigr)=(2,\,0.1,\,0)$
  and every subsequent value of $\gamma$ inherits the final state of the
  previous one at the same drive phase; the seed is reset at every
  column. In (a) the color scale is symmetric about $\lambda_{\max}=0$,
  chaotic points ($\lambda_{\max}>0$) appearing in violet to yellow and
  regular ones in gray to black. In (b) each point is classified as in
  Appendix~\ref{subsec:classification}, with the color code used throughout
  the paper: chaotic (black, two wells; violet and lilac, one well),
  antiperiodic (red), periodic and symmetry-broken (orange, two wells;
  yellow and pale yellow, one well); unclassified points (gray) are fewer
  than $0.1\%$ of the grid. The dashed lines mark the two one-parameter
  profiles studied below, $\gamma=1$ (Fig.~\ref{fig:fractions_omega}) and
  $\omega=1.3$ (Fig.~\ref{fig:fractions_gamma}), and the circles the
  sixteen parameter values at which the basins of attraction and the
  order maps are computed (Sec.~\ref{sec:basins}).}
\end{figure*}

Every trajectory is accordingly assigned to one of eight classes, which
combine the dynamical regime, chaotic, antiperiodic, or periodic but not
antiperiodic (i.e.\ symmetry-broken), with the wells visited, both, the
left one, or the right one, plus an unclassified class for the
trajectories whose regime cannot be resolved within the integration
window; the classification criteria are given in
Appendix~\ref{subsec:classification}, and the color code of the figures
is the same throughout [see Fig.~\ref{fig:plane_regimes}(b)].

Bifurcation diagrams constructed by continuation
(Appendix~\ref{subsec:continuation}) display an alternation of periodic,
antiperiodic, and chaotic windows, together with branches that appear or
disappear abruptly as the control parameter is varied, the
multistability-induced discontinuities, which signal that the trajectory
has jumped from the tracked attractor to a coexisting one. The symmetry
allows an antiperiodic window to end in only two ways: through a
supercritical pitchfork, at which the antiperiodic orbit sheds its
symmetry continuously and gives way to a conjugate pair of
symmetry-broken orbits of the same order, or through a saddle-node
bifurcation with an antiperiodic saddle, after which the trajectory
falls onto an unrelated attractor. One window of each kind, identified
in Ref.~I along the amplitude cut studied below, serves in
Sec.~\ref{sec:discussion} as the reference against which the basin
volumes are compared.

\section{Overview of the parameter plane}
\label{sec:plane}

Figure~\ref{fig:plane_regimes} sets the stage for what follows. It shows
the maximum Lyapunov exponent (a) and the dynamical regime (b) of the
attractor reached by numerical continuation along each vertical line of
the parameter plane (Appendix~\ref{subsec:continuation}). The region
$\omega\le1.6$ reproduces, at higher resolution, the maps of Ref.~I; the
extension to higher frequencies shows that above $\omega\simeq2.5$ the
response is a period-one orbit confined to one well at every amplitude
considered here. Three broad domains are apparent. At low amplitude,
and at all amplitudes for $\omega\gtrsim2.5$, the continued attractor is
a single-well periodic orbit, with a negative and featureless
$\lambda_{\max}$; its boundary is traced in panel~(a) by a narrow band
where $\lambda_{\max}$ approaches zero, the signature of the local
bifurcations through which the small orbit loses stability. Which of the
two conjugate wells appears in panel~(b) in such a region merely records
the member of the conjugate pair reached by the continuation and carries
no dynamical information.

Above this domain, for $\omega\lesssim1$, the plane is organized in
nearly vertical stripes in which two-well chaos, symmetry-broken
periodic orbits, and antiperiodic orbits alternate as the frequency is
varied at fixed amplitude, the antiperiodic stripes corresponding to
odd-order lockings of the drive. For $1\lesssim\omega\lesssim2.3$ a
large chaotic domain, crossed by oblique antiperiodic and periodic
bands, is bounded from below by a shrimp-shaped periodic structure,
located at $\omega\simeq1.0$--$1.3$, $\gamma\simeq0.25$--$0.45$, and
from the left by a broad antiperiodic domain, $\omega\simeq0.7$--$1.5$,
$\gamma\gtrsim0.5$, in which the symmetric orbit is the attractor
selected by the continuation over an extended parameter range.

Every point of Fig.~\ref{fig:plane_regimes} represents a single
attractor, so the map is blind to coexistence: wherever several
attractors coexist, it records the one selected by the continuation
history, and the sharp boundaries between colors that do not coincide
with a bifurcation of the followed orbit are the multistability-induced
discontinuities of Sec.~\ref{sec:model}. Since the oscillator is multistable over most
of this plane, the purpose of the present paper is to resolve the
coexistence behind this picture. We do so along two one-parameter cuts,
marked by the dashed lines in Fig.~\ref{fig:plane_regimes}, chosen to
traverse the two regions in which the map is richest.

The cut at fixed amplitude, $\gamma=1$, sweeps the low-frequency region,
where two-well chaos, symmetry-broken periodicity, and antiperiodicity
alternate in narrow stripes, crosses the broad antiperiodic domain and
the chaotic domain above it, and continues into the single-well regime
that sets in at $\omega\simeq2.5$. The cut at fixed frequency,
$\omega=1.3$, crosses, as $\gamma$ grows, the single-well regime, the
shrimp-shaped periodic structure, the chaotic domain with its
antiperiodic windows, and finally the broad antiperiodic domain. Along
each cut, eight parameter values, marked by circles in
Fig.~\ref{fig:plane_regimes}, are the points at which the basins of
attraction and the order maps are shown (Sec.~\ref{sec:basins}).

\section{Relative basin volumes along one-parameter profiles}
\label{sec:fractions}

A basin map on a two-dimensional section is an exact but partial
description: it fixes one of the three coordinates of the initial state,
and the areas it displays need not be representative of the volumes
occupied by the basins in the full space; in particular, conjugate
basins need not have equal areas on a section at fixed drive phase
(Sec.~\ref{sec:model}). To quantify how the coexisting regimes share the
space of initial conditions without singling out any section, we
estimate the relative volume $V_{c}$ of each basin within a fixed window
$W$ of the space of initial conditions by stochastic sampling, following
the approach introduced in Ref.~\cite{tarigo2024basin,tarigo2024basinMG} for delayed
systems, whose space of initial conditions is infinite-dimensional and
admits no exhaustive grid; in the terminology of
Ref.~\cite{menck2013basin} the quantity so obtained is the basin
stability of each attractor, resolved here as a function of a control
parameter and separated by dynamical regime, in the spirit of the Monte
Carlo basin bifurcation analysis of Ref.~\cite{gelbrecht2020monte} and
of the global continuation of Ref.~\cite{datseris2023framework}.

The profiles of $V_{c}$ against the forcing amplitude are also the
counterpart, for every coexisting regime at once, of the integrity
profiles introduced to quantify the decline of the safe basin of driven
oscillators as the excitation
grows~\cite{soliman1989integrity,rega2008dynamical}. The sample, of $4096$
initial conditions per parameter value, is stratified and built from
conjugate pairs, so that only half of it needs to be integrated and the
fractions of conjugate classes coincide exactly; the estimate is
unbiased, with a binomial standard error below $1.1\times10^{-2}$
(Appendix~\ref{subsec:sampling}).

\subsection{Amplitude profile at $\omega=1.3$}
\label{subsec:fractions_gamma}

\begin{figure*}[!t]
  \centering
  \includegraphics[width=0.85\textwidth]{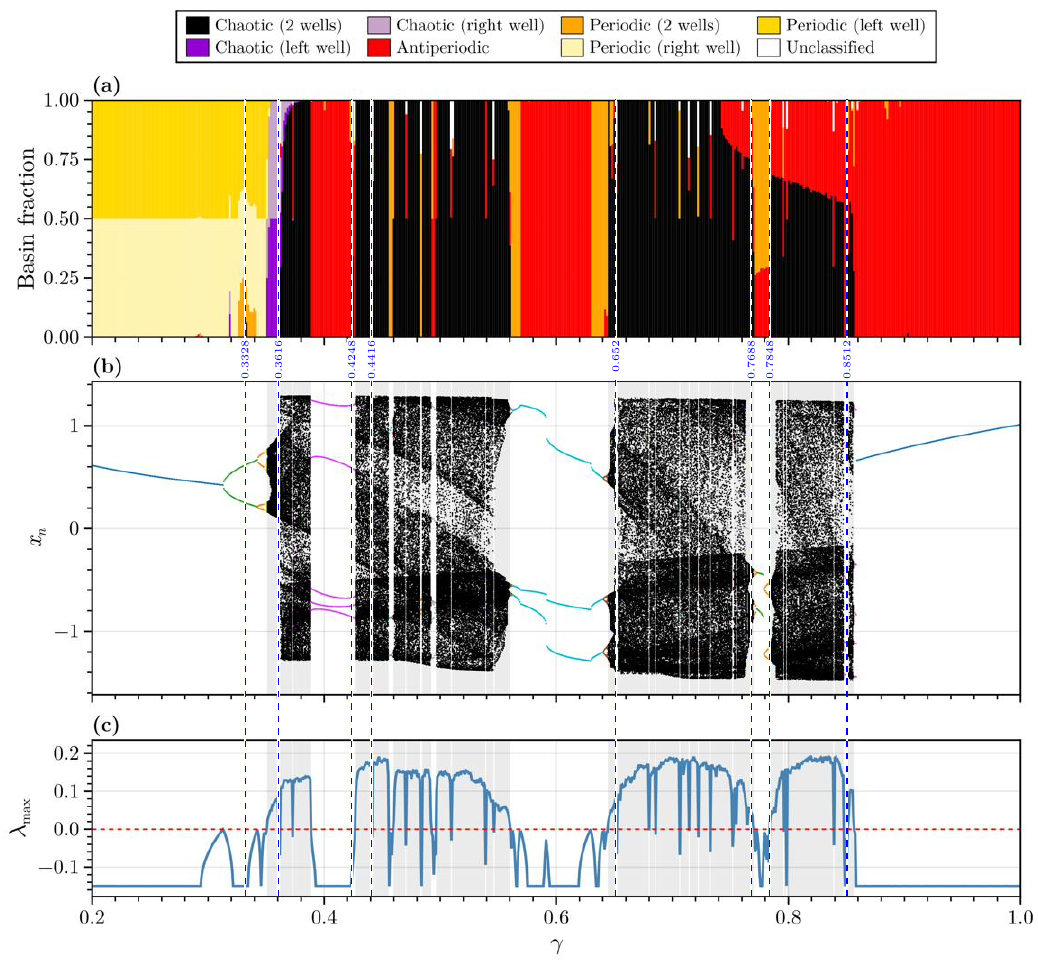}
  \caption{\label{fig:fractions_gamma}%
  Relative basin volumes and continuation sweep along the amplitude
  profile at $\omega=1.3$, on the same grid of $1001$ values of
  $\gamma\in[0.20,1.00]$ ($\Delta\gamma=8\times10^{-4}$).
  (a)~Fraction of initial conditions converging to each dynamical
  regime. At each value of $\gamma$, $4096$ initial conditions
  $(x(0),\dot{x}(0),\phi_{0})$ are drawn by $S$-symmetric stratified
  random sampling of the window $x(0)\in[-10,10]$, $\dot{x}(0)\in[-2,2]$,
  $\phi_{0}\in[0,2\pi)$ (Appendix~\ref{subsec:sampling}), integrated
  independently, and classified with the criteria of
  Appendix~\ref{subsec:classification}; the left- and right-well fractions
  coincide exactly by construction. Fractions are stacked so that each
  vertical bar adds up to unity; each bar corresponds to a single value
  of $\gamma$ ($N_{\mathrm{ind}}=2048$ independently integrated initial
  conditions, binomial standard error below $1.1\times10^{-2}$), and
  unclassified initial conditions appear as white gaps at the top of the
  stack. The color code of the legend is used in all the figures.
  (b)~Bifurcation diagram of a single trajectory followed by continuation
  over the same grid (Appendix~\ref{subsec:continuation}): stroboscopic values
  $x_{n}=x(nT_d)$ of the attractor reached at each $\gamma$, colored by
  the order $k$ of the orbit ($k=1$ blue, $2$ green, $3$ cyan, $4$ orange,
  $5$ magenta, \dots; black for chaotic). (c)~Maximum Lyapunov exponent
  of the same trajectory; the dashed red line marks $\lambda_{\max}=0$,
  and the gray bands in (b) and (c) mark the ranges in which the
  continuation is chaotic ($\lambda_{\max}>\varepsilon_{\lambda}$). The
  vertical dashed lines mark the eight values of $\gamma$ at which the
  basins of attraction and the order maps are shown in
  Figs.~\ref{fig:basins_gamma} and~\ref{fig:order_gamma}.}
\end{figure*}

Figure~\ref{fig:fractions_gamma}(a) shows the relative basin volumes
along the amplitude profile at $\omega=1.3$, stacked so that each
vertical bar adds up to unity. Panels (b) and (c) show, on the same grid of $1001$
values of $\gamma$, the bifurcation diagram and the maximum Lyapunov
exponent of a single trajectory followed by continuation
(Appendix~\ref{subsec:continuation}), colored by the order $k$ of the orbit
reached at each value; the vertical dashed lines mark the eight values
of $\gamma$ at which basin maps are shown in Sec.~\ref{sec:basins}.

Panel (a) condenses the multistability structure of the whole profile.
For $\gamma\lesssim0.32$ the window is split exactly in half between the
two conjugate single-well periodic orbits, of order $k=1$ up to the
period doubling at $\gamma\approx0.30$ visible in panel (b). The first
multistability window opens at $\gamma\approx0.325$--$0.345$, where a
symmetry-broken two-well periodic orbit captures up to $\approx30\%$ of
the window while coexisting with the single-well pair.

By $\gamma\approx0.35$ the single-well attractors have turned chaotic
and share the window exactly in half (fractions $0.50$ each), and at
$\gamma\approx0.36$ they merge into a single chaotic attractor visiting
both wells. This transition is a symmetry-increasing bifurcation in the
sense of Chossat and Golubitsky~\cite{chossat1988symmetry}, examined in
the driven double-well system in Ref.~\cite{ishii1986breakdown}. Since
two conjugate attractors are either identical or
disjoint~\cite{chossat1988symmetry,melbourne1993structure}, the attractor
born from the collision is necessarily $S$-invariant. Close to the
merger, trajectories linger in one well for long stretches before
hopping to the other, the crisis-induced intermittency characteristic of
such collisions~\cite{grebogi1987critical}. In the profile the merger
appears as a transition, spread over $\Delta\gamma\approx5\times10^{-3}$,
from the two conjugate single-well fractions to the two-well class,
during which more than a third of the sampled initial conditions remain
unclassified, their transients outlasting the classification window, and
an antiperiodic orbit of small basin ($V_{c}=0.12$ at $\gamma=0.3616$)
briefly takes part in the competition.

The antiperiodic windows stand out sharply: for
$0.39\lesssim\gamma\lesssim0.42$ and $0.57\lesssim\gamma\lesssim0.63$
the antiperiodic orbit, of order $k=5$ and $k=3$, respectively, as the
colors of panel (b) show, attracts essentially the entire window ($f=1$
within resolution over most of both ranges), so it is not merely present
but globally dominant there. These are the two windows of Ref.~I whose
ends serve as reference in Sec.~\ref{sec:model}: the lower one
terminates at the saddle-node bifurcation
$\gamma_{\mathrm{SN}}=0.4266645(2)$, the upper one at the supercritical
pitchfork $\gamma_{c}=0.629587$, beyond which the whole window passes to
the conjugate pair of two-well periodic orbits born there
($0.63\lesssim\gamma\lesssim0.644$).

Between and after these windows, in the chaotic bands $0.43\lesssim\gamma\lesssim0.56$ and
$0.645\lesssim\gamma\lesssim0.74$, the fractions change abruptly on the
scale of $\Delta\gamma$: periodic or antiperiodic attractors capture a
substantial part of the window over intervals of a few grid steps and
lose it just as quickly, while at the edges of every window at least two
classes coexist with comparable fractions.

For $0.745\lesssim\gamma\lesssim0.855$ the coexistence is no longer confined
to narrow windows: the antiperiodic basin grows gradually against
two-well chaos (from $0.18$ at $\gamma\approx0.75$ to $0.25$ at $0.767$,
and from $0.31$ at $\gamma\approx0.79$ to $0.44$ at $0.85$), interrupted
by a window ($0.769\lesssim\gamma\lesssim0.784$) in which a two-well
periodic orbit takes $\approx0.72$ of the window and the antiperiodic
orbit keeps $\approx0.27$. Finally, for $\gamma\gtrsim0.858$ and up to
the end of the range, an antiperiodic orbit of order $k=1$, the
large-amplitude symmetric response that straddles both wells, is the
unique attractor and captures every sampled initial condition.

The unclassified fraction remains below $0.05$ except in a few isolated bins,
located at the merger of the single-well chaotic attractors (maximum
$0.36$ at $\gamma=0.3616$) and inside the chaotic bands (up to
$\approx0.24$ near $\gamma\approx0.55$), where chaotic transients outlast
the classification window.

\subsection{Frequency profile at $\gamma=1$}
\label{subsec:fractions_omega}

\begin{figure*}[!t]
  \centering
  \includegraphics[width=0.85\textwidth]{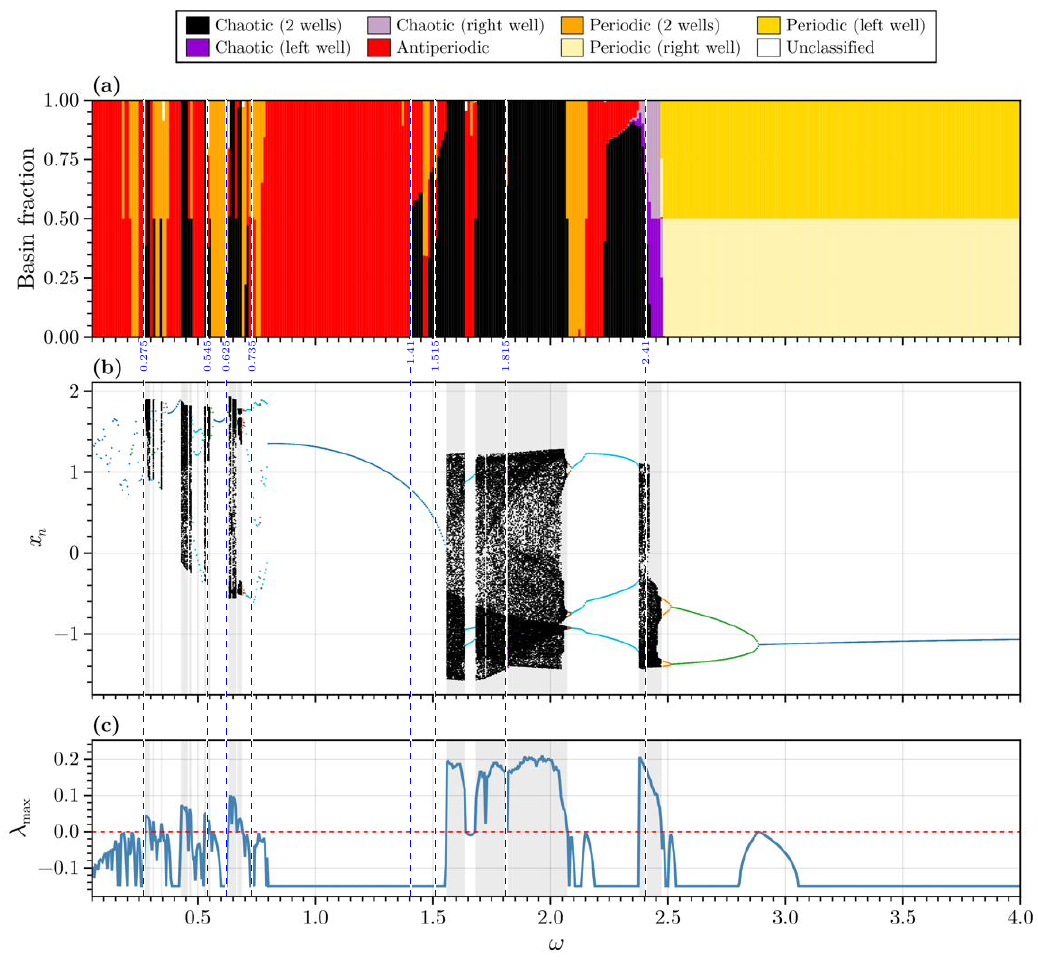}
  \caption{\label{fig:fractions_omega}%
  Relative basin volumes and continuation sweep along the frequency
  profile at $\gamma=1$, on the same grid of $791$ values of
  $\omega\in[0.05,4.00]$ ($\Delta\omega=5\times10^{-3}$). (a)~Fraction of
  initial conditions converging to each dynamical regime; sampling,
  classification, stacking, and color code follow the protocol of
  Fig.~\ref{fig:fractions_gamma}(a) (one bar per value of $\omega$,
  $N_{\mathrm{ind}}=2048$, binomial standard error below
  $1.1\times10^{-2}$). (b)~Stroboscopic values $x_{n}$ of the attractor
  reached by continuation at each $\omega$, colored by the order $k$, and
  (c)~its maximum Lyapunov exponent, as in
  Figs.~\ref{fig:fractions_gamma}(b) and (c). The vertical dashed lines
  mark the eight values of $\omega$ at which the basins of attraction and
  the order maps are shown in Figs.~\ref{fig:basins_omega}
  and~\ref{fig:order_omega}.}
\end{figure*}

Figure~\ref{fig:fractions_omega} applies the same stochastic sampling
and the same continuation at fixed $\gamma=1$, sweeping the drive
frequency over $\omega\in[0.05,4]$.

The frequency profile organizes into a sequence of distinct regimes.
Below $\omega\approx0.79$, antiperiodic dominance is repeatedly
interrupted: intervals where the antiperiodic orbit captures the entire
window (e.g., $0.05\leq\omega\lesssim0.18$, $0.38$--$0.43$, and
$0.48$--$0.53$) alternate with narrow bands dominated by two-well
periodic orbits (e.g., $0.22$--$0.25$, $0.56$--$0.61$, and $0.75$--$0.77$)
or by two-well chaos (e.g., $0.43$--$0.46$ and $0.64$--$0.69$), with
mixed bins at every boundary; the diagram of panel (b) shows the
corresponding succession of orbits of different order and of chaotic
bursts. For $0.79\lesssim\omega\lesssim1.40$ the antiperiodic orbit of
order $k=1$ attracts, within resolution, every sampled initial
condition: an uninterrupted plateau of global antiperiodic dominance.

At the ensemble level the loss of this plateau is gradual rather than
abrupt: between $\omega\approx1.40$ and $1.56$ the antiperiodic and
two-well chaotic attractors coexist while the antiperiodic basin
fraction shrinks progressively, from $0.42$ at $\omega\approx1.43$ to
$0.13$ at $1.55$, with a brief interlude near $\omega\approx1.48$ in
which a two-well periodic orbit takes a third of the window. Two-well
chaos then rules up to $\omega\approx2.07$, interrupted only by a narrow
antiperiodic window near $\omega\approx1.65$; a two-well periodic window
($2.08$--$2.15$) and a second high-frequency antiperiodic window
($2.16$--$2.23$, again with fraction one within resolution) follow,
before a final chaotic band ($2.25$--$2.42$) in which the antiperiodic
basin shrinks from $0.17$ to a few percent.

The exit from this band retraces, in reverse, the sequence found at low
amplitude along the $\gamma$ profile: at $\omega\approx2.43$ the
two-well chaotic attractor splits into a conjugate pair of single-well
chaotic attractors, which share the window exactly in half up to
$\omega\approx2.46$, and at $\omega\approx2.48$ the pair turns periodic
through an inverse period-doubling cascade, visible in panel (b), that
ends at $\omega\approx2.9$. For $\omega\gtrsim2.49$ and up to $\omega=4$
every sampled initial condition converges to one of the two conjugate
single-well periodic orbits (of order $k=2$ up to $\omega\approx2.9$ and
$k=1$ beyond), which split the window exactly in half. The unclassified
fraction remains below $0.1$ except in isolated bins located at window
boundaries, where chaotic transients outlast the classification window.

\section{Basins of attraction at the parameter values of strongest competition}
\label{sec:basins}

The profiles of Figs.~\ref{fig:fractions_gamma}(a)
and~\ref{fig:fractions_omega}(a) locate the parameter values at which
several attractors share the space of initial conditions, but they say
nothing about how they share it. To resolve the geometry of the
coexisting basins we select, on each profile, eight values at which the
competition is strongest (vertical dashed lines in
Figs.~\ref{fig:fractions_gamma} and~\ref{fig:fractions_omega}).
For each of these values two maps are shown on the section
$(x(0),\dot{x}(0))$ at drive phase $\phi_{0}=0$: the basin map, in which
every cell of the grid is colored by the class of the attractor it
converges to, and the order map, in which it is colored by the
stroboscopic order $k$ of that attractor.

The two maps are complementary. The basin map distinguishes the regimes
but lumps together all periodic attractors of the same well class, so
that, for instance, two coexisting two-well periodic orbits of different
order are indistinguishable in it. The order map separates them,
identifies the (odd) order of the antiperiodic attractors, and shows the
period-doubled state of the single-well orbits, at the price of not
distinguishing antiperiodic from symmetry-broken orbits of the same
order.

\subsection{Amplitude profile at $\omega=1.3$}
\label{subsec:basins_gamma}

\begin{figure*}[!t]
  \centering
  \includegraphics[width=\textwidth]{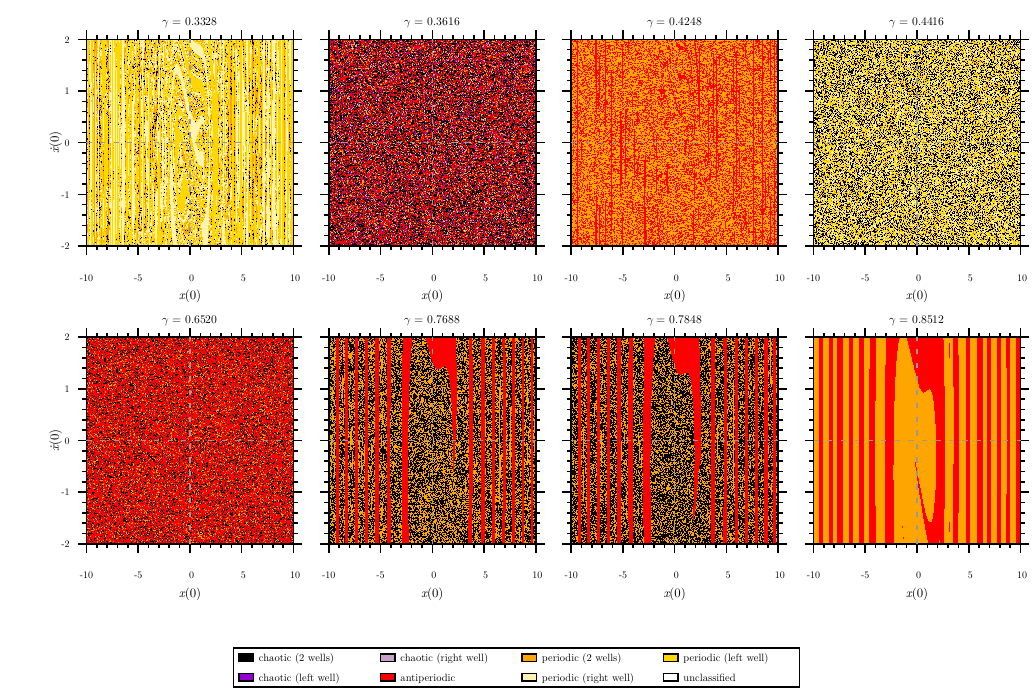}
  \caption{\label{fig:basins_gamma}%
  Basins of attraction on the section $(x(0),\dot{x}(0))$ of the space of
  initial conditions, at drive phase $\phi_{0}=0$ and fixed $\omega=1.3$,
  for the eight values of the forcing amplitude marked in
  Fig.~\ref{fig:fractions_gamma} (the value of $\gamma$ is indicated
  above each panel). Each initial condition of an $800\times320$ grid spanning
  $x(0)\in[-10,10]$ and $\dot{x}(0)\in[-2,2]$ is integrated independently
  and classified according to its asymptotic regime
  (Appendix~\ref{subsec:basins}); the color code is that of
  Fig.~\ref{fig:fractions_gamma}, repeated in the legend at the bottom,
  with unclassified initial conditions in white. Dashed lines mark
  $x(0)=0$ and $\dot{x}(0)=0$.}
\end{figure*}

\begin{figure*}[!t]
  \centering
  \includegraphics[width=\textwidth]{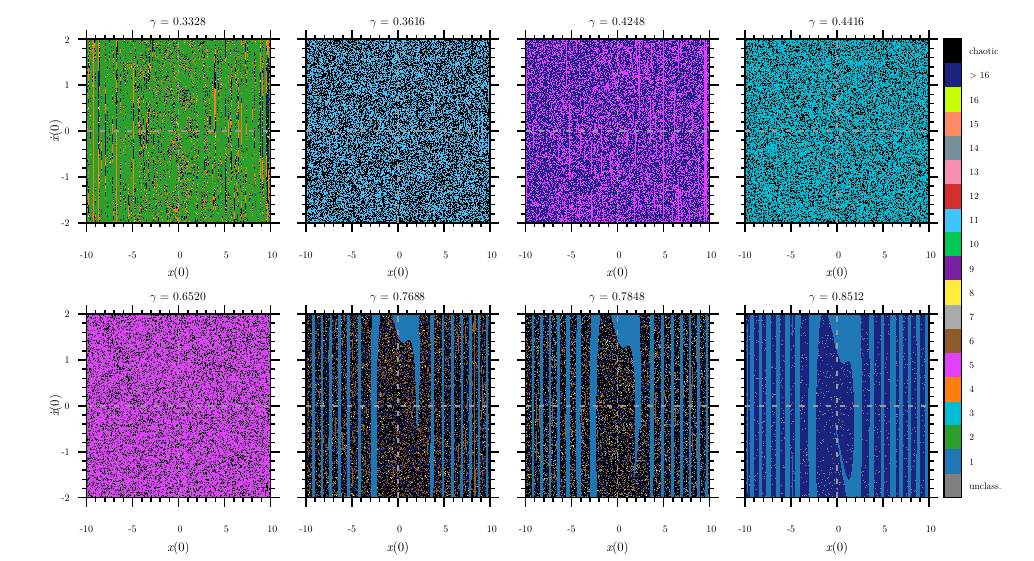}
  \caption{\label{fig:order_gamma}%
  Order $k=T_{\mathrm{orb}}/T_d$ of the attractor reached from each
  initial condition of Fig.~\ref{fig:basins_gamma}, on the same sections
  and for the same eight values of $\gamma$. The categorical colorbar
  distinguishes the orders $k=1,\dots,16$ ($k=1$ blue), collects the
  orders above $16$ in a single color, and reserves black for chaotic
  attractors and white for unclassified initial conditions. Read together
  with Fig.~\ref{fig:basins_gamma}, these maps separate coexisting
  periodic attractors of different order that fall in the same class of
  the regime classification, and identify the odd order of the
  antiperiodic attractors.}
\end{figure*}

Figure~\ref{fig:basins_gamma} shows the basins of attraction at the
eight selected values of $\gamma$, and Fig.~\ref{fig:order_gamma} the
corresponding order maps. The eight panels illustrate qualitatively distinct forms of
multistability, in the order in which they are met along the profile.
At $\gamma=0.3328$ three periodic attractors coexist: the conjugate pair
of single-well orbits, in their period-doubled state, which covers
$\approx80\%$ of the section, and a symmetry-broken two-well periodic
orbit of higher order, whose basin occupies $\approx12\%$ of it. At
$\gamma=0.3616$ the single-well attractors have become chaotic and the
map is taken at their merger into a two-well chaotic attractor: the
conjugate pair of single-well chaotic attractors ($\approx12\%$ of the
section) coexists with two-well chaos ($\approx37\%$) and with an
antiperiodic orbit whose basin covers $\approx43\%$ of the section but
only $12\%$ of the window, while more than a third of the sampled initial
conditions of the window remain unclassified --- the signature, at the
level of the basins, of the symmetry-increasing crisis at which the two
conjugate basins coalesce into a single $S$-invariant one and chaotic
transients become long.

At $\gamma=0.4248$ the antiperiodic orbit of order $k=5$, less than
$2\times10^{-3}$ below its saddle-node termination at
$\gamma_{\mathrm{SN}}$ (Sec.~\ref{subsec:fractions_gamma}), has already
lost most of the section ($\approx39\%$) to a symmetry-broken two-well
periodic orbit of higher order ($\approx61\%$); the order map identifies
both. At $\gamma=0.4416$, inside the chaotic band that follows the fold,
a conjugate pair of single-well periodic orbits captures $\approx62\%$
of the section, coexisting with two-well chaos ($\approx26\%$) and with
a two-well periodic orbit of small basin --- one of the narrow windows
that interrupt the band in Fig.~\ref{fig:fractions_gamma}(a). At
$\gamma=0.6520$, at the boundary between the periodic window born at the
pitchfork and the subsequent chaotic band, an antiperiodic orbit covers
$\approx73\%$ of the section while two-well chaos and a two-well periodic
orbit retain $\approx18\%$ and $\approx7\%$; two further values of the
chaotic bands, $\gamma=0.5464$ and $0.7144$, display the same
antiperiodic--chaotic coexistence and are not shown.

The last three
panels belong to the range of persistent coexistence: at $\gamma=0.7688$
and $\gamma=0.7848$, at the entrance and at the exit of the two-well
periodic window $0.769$--$0.784$, chaotic, antiperiodic, and two-well
periodic attractors coexist with basins of comparable area
($35/29/25\%$ and $37/34/19\%$ of the section, respectively); and at
$\gamma=0.8512$, at the end of the last chaotic band, the two-well
periodic and antiperiodic orbits that are about to replace it already
share the whole section ($58\%$ and $42\%$), the chaotic attractor
retaining $15\%$ of the window but no visible basin on this section. In
all panels the basins are intertwined down to the resolution of the grid
over extended regions, the fractal basin boundaries characteristic of
this system~\cite{moon1985fractal,aguirre2002unpredictable,puy2021test},
while compact cores surround the attractors themselves.

\subsection{Frequency profile at $\gamma=1$}
\label{subsec:basins_omega}

\begin{figure*}[!t]
  \centering
  \includegraphics[width=\textwidth]{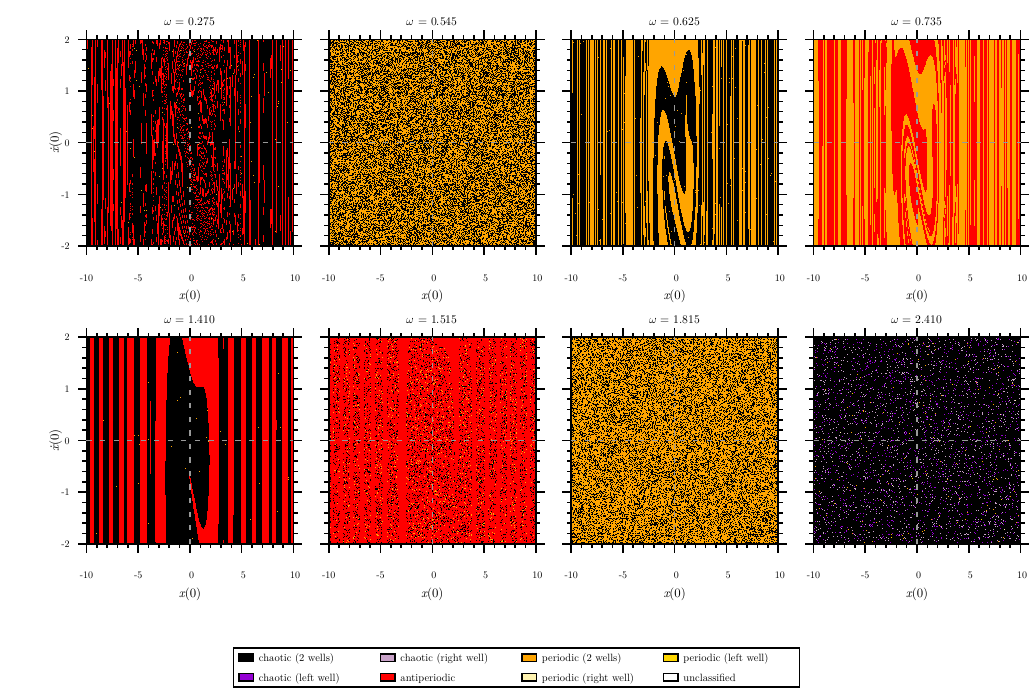}
  \caption{\label{fig:basins_omega}%
  Basins of attraction on the section $(x(0),\dot{x}(0))$ at drive phase
  $\phi_{0}=0$ and fixed $\gamma=1$, for the eight values of the drive
  frequency marked in Fig.~\ref{fig:fractions_omega} (the value of
  $\omega$ is indicated above each panel). Grid, integration, classification, color code, and
  reference lines as in Fig.~\ref{fig:basins_gamma}.}
\end{figure*}

\begin{figure*}[!t]
  \centering
  \includegraphics[width=\textwidth]{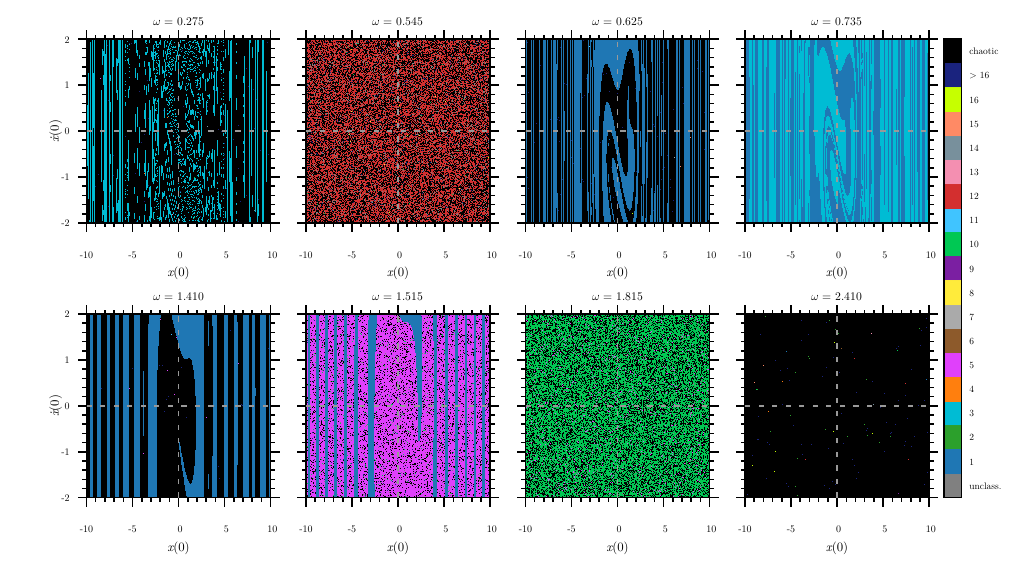}
  \caption{\label{fig:order_omega}%
  Order $k$ of the attractor reached from each initial condition of
  Fig.~\ref{fig:basins_omega}, on the same sections and for the same
  eight values of $\omega$. Color code as in Fig.~\ref{fig:order_gamma}.}
\end{figure*}

Figures~\ref{fig:basins_omega} and~\ref{fig:order_omega} show the
basins of attraction and the order maps at the eight selected values of
$\omega$.
At low frequency the panels sample the dense alternation of windows of
Fig.~\ref{fig:fractions_omega}(a). At $\omega=0.275$, at the boundary
between an antiperiodic window and a chaotic band, two-well chaos
dominates the section ($\approx80\%$) and an antiperiodic orbit of high
order retains the rest. At $\omega=0.545$, at the entrance of the
two-well periodic window $0.56$--$0.61$, a two-well periodic orbit
($\approx48\%$) and two-well chaos ($\approx36\%$) share the section,
with a noticeable set of unclassified cells at their boundaries, and at
$\omega=0.625$, at the exit of the same window, two-well chaos
($\approx53\%$) and the two-well periodic orbit ($\approx44\%$) coexist
in nearly equal parts. At $\omega=0.735$ an antiperiodic orbit and a
symmetry-broken two-well periodic orbit share both the section and the
window exactly in half, with no chaotic component.

The next two panels document the end of the plateau of global
antiperiodic dominance: at $\omega=1.41$ the antiperiodic orbit of order
$k=1$, which captured every initial condition up to $\omega\approx1.40$,
coexists with the two-well chaotic attractor that has just appeared, in
comparable proportions both on the section ($45\%$) and in the window
($V_{c}=0.46$); at $\omega=1.515$, inside the interval of gradual
decline, the antiperiodic orbit still covers $\approx91\%$ of the
section, although its volume in the window has shrunk to less than a
third; the loss of its basin, at this frequency, takes place at drive
phases other than $\phi_{0}=0$. At $\omega=1.815$, inside the chaotic
band $1.69$--$2.07$, a two-well periodic orbit captures $\approx61\%$ of
the section and coexists with two-well chaos.

At $\omega=2.41$, finally, the map is taken at the end of the last
chaotic band, where the two-well chaotic attractor coexists with the
conjugate pair of single-well chaotic attractors into which it splits at
$\omega\approx2.43$ --- the symmetry-increasing crisis of
Fig.~\ref{fig:basins_gamma} ($\gamma=0.3616$), traversed in the opposite
direction; here the two-well attractor covers $\approx95\%$ of the
section while the window is split evenly between it ($V_{c}=0.51$) and
the single-well pair ($0.25$ each).

The last two examples show that the areas on a section at fixed drive
phase can differ markedly from the volumes in the window, in either
direction. The reason is the symmetry (Sec.~\ref{sec:model}): the
basins of two conjugate attractors have exactly equal volumes in the
full space of initial conditions, and hence exactly equal fractions in
Figs.~\ref{fig:fractions_gamma}(a) and~\ref{fig:fractions_omega}(a), but
need not have equal areas on a section at fixed drive phase, on which
they may look quite different [Figs.~\ref{fig:basins_gamma}
and~\ref{fig:basins_omega}, $\gamma=0.3328$ and $\omega=2.41$]; nor do the basins on these sections
display, in general, the point symmetry
$(x(0),\dot{x}(0))\mapsto(-x(0),-\dot{x}(0))$.

On sections $(x(0),\phi_{0})$ at fixed $\dot{x}(0)=0$ (not shown), on
which $S$ acts as the half-turn
$(x(0),\phi_{0})\mapsto(-x(0),\phi_{0}+\pi)$, the conjugacy between twin
basins was verified cell by cell at all the selected values: $99.9\%$ of
the cells are assigned to the class conjugate to that of their image
under $S$, the exceptions lying on basin boundaries, where the finite
transient cannot resolve the asymptotic regime of an individual initial
condition.

The drive phase is thus a coordinate of the initial state that cannot be
ignored in the description of the multistability of a driven system. In
particular, the attractor reached in a parameter sweep is selected also
by the phase at which each parameter value is sampled; this underlies
the multistability-induced discontinuities of the continuation diagrams,
discussed next.

\section{Discussion: basin volumes versus continuation sweeps}
\label{sec:discussion}

The bifurcation diagrams of Figs.~\ref{fig:fractions_gamma}(b)
and~\ref{fig:fractions_omega}(b) and the basin-volume profiles of
Figs.~\ref{fig:fractions_gamma}(a) and~\ref{fig:fractions_omega}(a)
describe the same two one-parameter cuts of the $(\omega,\gamma)$ plane,
on the same grids and with the same classifier, from opposite
standpoints. A continuation sweep follows one trajectory: at each
parameter step the final state of the previous value is used as the
initial condition, and the sweep remains on a branch as long as the
corresponding attractor exists, is stable, and the inherited state
(displaced by the parameter change) still lies in its basin. The
stochastic sampling, in contrast, weighs every attractor by the volume
of its basin, regardless of history.

Three kinds of events therefore
leave different signatures in the two descriptions: a bifurcation of the
tracked attractor, which appears in both; a jump of the sweep to a
coexisting attractor, which appears in the diagram as a discontinuity
and in the profile as a coexistence of classes with substantial
fractions; and a coexistence that the sweep never visits, which appears
in the profile only. Since the class of the continuation is available at
every grid value, the comparison can be made quantitative: the
continuation lies in the class that dominates the sample at most values
of both profiles, and in a minority class ($f<0.5$) only at a small
number of values, all of them adjacent to a discontinuity of the
diagram.
We now go through the two profiles with this dictionary in hand.

\subsection{Amplitude profile at $\omega=1.3$}

For $\gamma\lesssim0.32$ the profile shows the two conjugate single-well
orbits sharing the window exactly in half, whereas the diagram displays
a single branch, confined to the well selected by the reference initial
state of the sweep. This is the simplest instance of a general fact that
follows from the symmetry alone (Sec.~\ref{sec:model}): every single-well
regime is bistable by construction, and which of the two twins is
realized is a matter of history rather than of dynamics. The same fact
explains an observation made in Ref.~I for the two-parameter maps,
namely that the regime classification of the $(\omega,\gamma)$ plane is
insensitive to the sweep strategy while the well selected by single-well
orbits is not: the regime statistics count both twins alike, but the
twin selected at each grid point depends on the direction of the sweep.

The first multistability window, $\gamma\approx0.325$--$0.345$, in which
a symmetry-broken two-well periodic orbit captures up to $\approx30\%$ of
the window, is an example of a coexistence that a continuation sweep
starting from a single-well state does not visit: the sweep of
Fig.~\ref{fig:fractions_gamma}(b) remains on the single-well branch,
which retains the larger basin, and the two-well orbit is invisible to
it although it attracts almost a third of the initial conditions
[Fig.~\ref{fig:basins_gamma}, $\gamma=0.3328$].

The subsequent conversion of the single-well orbits into single-well
chaos, near $\gamma\approx0.35$, and the merger of the two conjugate
chaotic attractors into a two-well one near $\gamma\approx0.36$
[Fig.~\ref{fig:basins_gamma}, $\gamma=0.3616$], appear in both
descriptions in the same way, as they involve no jump: the chaotic band
of the diagram widens from one sign of $x$ to both wells at the
symmetry-increasing crisis, and the two conjugate fractions of $0.50$ of
the profile coalesce into the two-well chaotic class.

The two antiperiodic windows of Sec.~\ref{subsec:fractions_gamma}
provide the sharpest contrast between the two descriptions. The window
of order $k=5$ ends, for the continuation, at the saddle-node
bifurcation $\gamma_{\mathrm{SN}}=0.4266645$, where the antiperiodic orbit
is annihilated and the trajectory falls onto the chaotic attractor that
fills the band beyond it. At the ensemble level, however, the
antiperiodic orbit ceases to be globally dominant earlier, at
$\gamma\approx0.422$: over the last $\Delta\gamma\approx5\times10^{-3}$
of the window it coexists with a symmetry-broken periodic orbit, which
takes between a sixth and, at $\gamma=0.4248$, more than half of the
window [$V_{c}=0.57$, Fig.~\ref{fig:basins_gamma}, $\gamma=0.4248$], until
the fold removes the antiperiodic orbit altogether.

The loss of the basin thus precedes the loss of stability, and the
window has two different upper ends according to whether it is defined
by the existence of the attractor or by its dominance; a continuation
sweep, which stays on the branch until it disappears, necessarily
overshoots the range of dominance. The reorganization of coexisting
basins around a saddle-node bifurcation, at which a basin is created or
destroyed, was described for driven two-dimensional systems in
Ref.~\cite{eschenazi1989basins}, and the sharp decline of the basin of a
driven oscillator ahead of the fold at which the attractor itself
disappears is a well-documented mechanism, both
numerically~\cite{soliman1992global} and
experimentally~\cite{todd1997experimental}.

The window of order $k=3$ behaves in the opposite way. The antiperiodic
orbit attracts the entire window up to the pitchfork at
$\gamma_{c}=0.629587$, and beyond $\gamma_{c}$ the whole window passes to
the two-well periodic class, i.e.\ to the conjugate pair of
symmetry-broken orbits of the same order born at the bifurcation, with
no other class in between. This is what a supercritical pitchfork
requires: the antiperiodic orbit becomes unstable at $\gamma_{c}$ and its
basin is inherited by the two twins, split into two conjugate halves; no
third attractor competes, no jump takes place, and the end of the window
is the same point in both descriptions.

The two exits allowed by the symmetry for an $S$-invariant orbit are
therefore distinguished not only by the local signatures established in
Ref.~I --- the continuous growth of the residual $\eta$ and the linear
crossing of $\lambda_{\max}$ at the pitchfork, against the discontinuous
jump of $\eta$ and the square-root vanishing of $\lambda_{\max}$ at the
fold --- but also by the fate of the basin: continuous handover to the
conjugate pair at the pitchfork, competition with an unrelated attractor
followed by a jump at the fold.

Inside the chaotic bands, and for $0.745\lesssim\gamma\lesssim0.855$,
the profile displays abrupt changes of the fractions on the scale of
$\Delta\gamma$, reminiscent of the basin explosions of
Ref.~\cite{ueda1990basin}, and extended coexistence of two or three
classes in comparable proportions [Fig.~\ref{fig:basins_gamma},
$\gamma=0.4416$ to $0.8512$]. These are
the regions in which the diagram shows branches that appear and
disappear abruptly. The correspondence is one-way: wherever the
continuation jumps between branches, the profile shows two attractors
with substantial basins, but a coexistence may go unnoticed by the sweep
as long as the inherited state remains in the basin of the tracked
attractor.

Since the basins of Sec.~\ref{sec:basins} are intertwined down to the
resolution of the grid over extended regions of the section, since
basin boundaries may themselves change discontinuously as a parameter
is varied~\cite{grebogi1986metamorphoses}, and since they depend on the
drive phase $\phi_{0}$ at which the state is handed over, the
displacement of the inherited state by the parameter step acts as a
finite kick, which suffices to change basin wherever the boundaries are
intertwined; a continuation carried out coherently with the forcing, as
here, fixes the phase but not the kick. The range of global dominance of
the $k=1$ antiperiodic orbit for $\gamma\gtrsim0.858$ is, in contrast, a
range in which the two descriptions necessarily agree: where a single
attractor exists there is nothing to jump to.

\subsection{Frequency profile at $\gamma=1$}

The frequency profile confirms the same picture along the other cut. The
plateau of global antiperiodic dominance, $0.79\lesssim\omega\lesssim1.40$,
coincides with an interval over which the diagram follows the
antiperiodic branch of order $k=1$ without discontinuities. The end of
the plateau illustrates once more the difference between the loss of
dominance and the loss of the branch. Between $\omega\approx1.40$ and
$1.56$ the antiperiodic and two-well chaotic attractors coexist and the
antiperiodic fraction declines gradually, from $0.42$ at $\omega\approx1.43$
to $0.13$ at $1.55$ [Fig.~\ref{fig:basins_omega}, $\omega=1.41$ and
$1.515$]; the sweep, in contrast, abandons the antiperiodic branch at a
single value of $\omega$, near the end of this interval.

The section at $\omega=1.515$ shows why the phase matters: on the
section $\phi_{0}=0$ the antiperiodic orbit still holds $\approx91\%$ of
the initial conditions, whereas in the full window its basin has shrunk
to less than a third, so that a sweep handed over at a different phase
would already have left the branch. The frequency at which the sweep
leaves the branch thus depends on the reference initial state and on
the drive phase, and may fall anywhere within the interval of decline;
the ensemble description replaces this history-dependent value by a
smooth and intrinsic function, the antiperiodic basin fraction.

At low frequencies, $\omega\lesssim0.79$, the dense alternation of narrow
windows in the profile, with mixed bins at every boundary, matches the
dense succession of transitions in the diagram along the same sweep. The
mixed bins show that the transitions are generally accompanied by
coexistence [Fig.~\ref{fig:basins_omega}, $\omega=0.275$ to $0.735$], so
that the position of each boundary in the diagram is subject to the same
history dependence.

At high frequencies the exit from the last chaotic band retraces in
reverse the sequence found at low amplitude along the $\gamma$ profile
(two-well chaos, conjugate single-well chaos, conjugate single-well
periodic orbits) and appears in both descriptions in the same way, since
no jump is involved: the diagram shows the chaotic band narrowing to one
sign of $x$ and then an inverse period-doubling cascade of the
single-well orbit, and the profile shows the two-well class giving way
to two conjugate fractions of $0.50$ [Fig.~\ref{fig:basins_omega},
$\omega=2.41$, where the section, unlike the window, is still almost
entirely two-well chaotic].

For $\omega\gtrsim2.49$ the window is split exactly in half between the
two conjugate single-well periodic orbits, of which the diagram shows
one. The antiperiodic orbit essentially never coexists with this pair
anywhere along the profile, so that the transition from antiperiodic to
single-well dynamics always proceeds through intermediate chaotic or
two-well periodic regimes.

\subsection{General remarks}

Three conclusions of general scope emerge from the comparison. First, the
multistability-induced discontinuities of continuation sweeps are
located precisely where the basin volumes change abruptly, and each of
them corresponds to a coexistence of attractors with substantial basins;
the converse does not hold, since a coexisting attractor whose basin
does not contain the inherited state is invisible to the sweep; basin
transformations that leave no trace in a bifurcation diagram have
likewise been detected by following the basin entropy along a
parameter~\cite{wagemakers2023using}. Second,
the boundary of a regime window has two meanings, the existence and
stability of the attractor and its dominance over the space of initial
conditions, which coincide when the window ends through a bifurcation of
the attractor alone (the pitchfork) but differ when it ends through a
collision that leaves a coexisting attractor to take over (the fold).
Third, and specific to systems with a symmetry that exchanges attractors,
symmetry-broken attractors always come in pairs with basins of exactly
equal volume, so that any regime of broken symmetry is at least bistable
by construction, whereas an $S$-invariant attractor can be, and over
extended parameter ranges is, the unique attractor of the system: global
dominance is a privilege of the symmetric attractors.

\section{Conclusions}
\label{sec:conclusions}
We have characterized the multistability of the periodically driven
Duffing--Holmes oscillator in the space of initial conditions
$(x(0),\dot{x}(0),\phi_{0})$, using two tools. Stratified random
sampling of the full three-dimensional space yields an unbiased estimate
of the relative volume of every basin, with binomial error bars, as a
function of a control parameter and on the same grid on which the
bifurcation diagrams of a single trajectory are computed by
continuation. Basin maps on two-dimensional sections, together with maps
of the order of the attractor reached from each initial condition,
resolve the geometry of the coexisting basins where the competition
between attractors is strongest. The sampling exploits the half-period
shift symmetry $S$ twice. Since the window and the sampling measure are
$S$-invariant, only half of the sample needs to be integrated, and the
fractions of conjugate classes coincide exactly by construction.

The basin maps show that the Duffing--Holmes oscillator realizes
qualitatively distinct forms of multistability at nearby values of the
forcing parameters: several periodic orbits; single-well and two-well
chaos with an antiperiodic orbit, at a symmetry-increasing crisis; an
antiperiodic orbit with a symmetry-broken periodic orbit, at the edge of
its window; and periodic, antiperiodic, and chaotic attractors together.
The order maps separate coexisting periodic attractors of different
order, which the regime classification lumps together, and identify the
odd order of the antiperiodic ones. Since $S$ maps a section at fixed
drive phase onto the section at the opposite phase, conjugate basins
have exactly equal volumes in the full space but not equal areas on a
section. The area of a basin on a section may therefore differ markedly
from its relative volume in the window, in either direction. The drive
phase $\phi_{0}$ is thus a coordinate of the initial state that cannot
be ignored in the description of the multistability of a driven system.

The relative basin volumes along the amplitude profile at $\omega=1.3$
and the frequency profile at $\gamma=1$ reveal windows of global
antiperiodic dominance, in which the antiperiodic orbit attracts
essentially every initial condition. These windows are separated by
narrow intervals in which two or three classes share the space of
initial conditions in comparable proportions, and an extended range of
persistent coexistence appears for $0.745\lesssim\gamma\lesssim0.855$.
Compared with the continuation sweeps on the same grids, the abrupt
changes of basin volume are the ensemble-level counterpart of the
multistability-induced discontinuities. The two ways in which an
antiperiodic window can end acquire distinct ensemble signatures: a
continuous handover of the basin to the conjugate pair at the
pitchfork, and a competition with an unrelated attractor followed by a
jump at the fold. Finally, the gradual shrinking of the antiperiodic
basin at the end of a window of dominance explains why the frequency at
which a sweep abandons the antiperiodic branch is a property of the
sweep, not of the system.
\begin{acknowledgments}
A.C.M.\ acknowledges financial support from Project CSIC I+D \textit{Predictabilidad, caos, regularidad y simetr\'ias en sistemas f\'isicos no lineales}, funded by CSIC--UdelaR.
E.D.L. acknowledges support from the Brazilian agencies CNPq (No. 304398/2023-3) and FAPESP (2025/14544-0). The authors acknowledge computing time on the high-performance cluster \texttt{ClusterUY}.
\end{acknowledgments}

\appendix

\section{Numerical methods}
\label{app:methods}

\subsection{Classification of individual trajectories}
\label{subsec:classification}

All computations were carried out in Julia with the
\textsc{DynamicalSystems.jl} ecosystem~\cite{Datseris2018}. Unless
otherwise stated, the values that follow are used throughout.
Equation~\eqref{eq:duffing} is integrated with the ninth-order explicit
Runge--Kutta scheme of Verner (\texttt{Vern9}) under adaptive step-size
control, with absolute and relative tolerances of $10^{-9}$. For each
initial condition a transient of duration $T_{\mathrm{tr}}=500$ is
discarded and the asymptotic trajectory is recorded over a window
$T_{\mathrm{int}}=1.4\times10^{4}$. The maximum Lyapunov exponent
$\lambda_{\max}$ is computed from the variational equations over a time
$T_{\lambda}=10^{4}$ on the continuation of the recorded trajectory,
i.e.\ from its final state and at the corresponding drive phase, so that
the exponent refers to the attractor that has been classified.

A trajectory is chaotic if $\lambda_{\max}>\varepsilon_{\lambda}=10^{-3}$.
Otherwise, its order $k$ is obtained from the stroboscopic section
$x_{n}=x(nT_d)$ as the smallest lag for which the section recurs to
within a tolerance $\varepsilon_{r}=10^{-2}$ relative to the size of the
orbit. Antiperiodicity is quantified by the amplitude-normalized
residual
\begin{equation}
  \eta=\frac{1}{A}\,
  \Big\langle\,\big|\,x(t)+x\!\left(t+\tfrac{1}{2}T_{\mathrm{orb}}\right)\big|\,\Big\rangle ,
  \label{eq:antip_residual}
\end{equation}
where $A$ is the orbit amplitude and the average runs over the recorded
window; a periodic orbit is antiperiodic if
$\eta<\varepsilon_{a}=10^{-2}$. The wells visited are determined from
the sign of $x$ over the last $T_{\mathrm{w}}=100$ time units of the
window. Non-chaotic trajectories for which no stroboscopic recurrence is
detected are left unclassified.

The continuation sweeps and the basin computations differ not in the
classifier but in the choice of initial conditions: in a sweep each
parameter value inherits the final state of the previous one
(Appendix~\ref{subsec:continuation}), whereas in the basin maps and in
the basin-volume estimates every initial condition, of a fixed grid or
of the random sample, is integrated independently from the prescribed
state $(x(0),\dot{x}(0),\phi_{0})$, so that the regime assigned to it
does not depend on the order in which the points or the parameter values
are traversed.

\subsection{Basin maps on two-dimensional sections}
\label{subsec:basins}

The basin maps of Sec.~\ref{sec:basins} are computed on the section
$\phi_{0}=0$ of the space of initial conditions, on an $800\times320$
grid spanning $x(0)\in[-10,10]$ and $\dot{x}(0)\in[-2,2]$, at the sixteen
parameter values of Figs.~\ref{fig:basins_gamma}
and~\ref{fig:basins_omega}.
Trajectories are integrated with
the fifth-order scheme of Tsitouras (\texttt{Tsit5}), with the same
tolerances, and sampled at 200 points per drive period. After a transient
of $400\,T_d$ the orbit is classified over a window of $160\,T_d$: a
stroboscopic period $p\le64$ is accepted when the last eight section
points recur to within $10^{-3}A$, with $A=\max|x|$, and the orbit is
flagged as antiperiodic when $|x(t+pT_d/2)+x(t)|<0.05\,A$ pointwise over
the second half of the window.

If neither test closes, the transient is extended by $1200\,T_d$, up to
twice, so that chaotic supertransients can decay, and only then is
$\lambda_{\max}$ estimated from the divergence of two nearby
trajectories over $1500\,T_d$; the orbit is labeled chaotic if
$\lambda_{\max}>10^{-2}$, and regular orbits whose period exceeds
$64\,T_d$ are counted as periodic. An orbit is assigned to both wells
when $x$ exceeds $\pm0.05$ on both sides within the window. These looser
thresholds are adequate for delineating basins away from bifurcation
points, where the coexisting regimes are well separated; the basin
volumes of Sec.~\ref{sec:fractions} rely instead on the full pipeline of
Appendix~\ref{subsec:classification}. For every cell of the grid the
stroboscopic order $k$ of the attractor reached is recorded as well, and
the area $A_{c}$ occupied by each class on the section is obtained by
counting cells.

\subsection{Stochastic estimation of relative basin volumes}
\label{subsec:sampling}

Let $W$ denote the window
$x(0)\in[-10,10]$, $\dot{x}(0)\in[-2,2]$, $\phi_{0}\in[0,2\pi)$ of the
space of initial conditions, which contains both wells, all the
attractors found in the parameter ranges considered, and a wide margin
around them. For a class $c$ of trajectories with basin
$\mathcal{B}_{c}$, the relative basin volume is
\begin{equation}
  V_{c}=\frac{\mathrm{vol}(\mathcal{B}_{c}\cap W)}{\mathrm{vol}(W)},
  \label{eq:volume}
\end{equation}
i.e.\ the probability that an initial condition drawn uniformly from $W$
converges to an attractor of class $c$; the volumes of all classes,
including the unclassified one, add up to unity.

At each value of the control parameter, $V_{c}$ is estimated by the
fraction $f_{c}$ of a random sample of $N$ initial conditions assigned
to class $c$. The sample is stratified: each axis of $W$ is divided into
$16$ equal intervals, which partition $W$ into $16^{3}=4096$ cells of
equal volume, and one initial condition is drawn uniformly from each
cell, so that $N=4096$. Stratification preserves the uniform sampling
measure, and hence the unbiasedness of $f_{c}$, while it suppresses the
clustering of a purely random sample and reduces its variance below the
binomial value. The pseudo-random generator is seeded deterministically
from the value of the control parameter, so that every run is
reproducible and the samples at different parameter values are
statistically independent; a feature that persists over several
consecutive values is therefore not a sampling artifact.

Both the window $W$ and the stratified sampling measure are invariant
under the action of $S$ on initial conditions: the intervals in $x(0)$
and $\dot{x}(0)$ are symmetric about the origin, and the shift
$\phi_{0}\mapsto\phi_{0}+\pi$ maps the partition of $[0,2\pi)$ onto
itself. The sample is therefore built as a set of conjugate pairs
$\{\mathbf{p},S(\mathbf{p})\}$: the $2048$ representatives are drawn from
the cells with $\phi_{0}\in[0,\pi)$ and their images under $S$ complete
the sample. Only the representatives need to be integrated, since the
class of the image point follows from that of the representative by
exchanging the wells ($\lambda_{\max}$, $k$, and $\eta$ being invariant
under $S$); the cost is halved, and the left- and right-well fractions
coincide exactly by construction.

The fraction $f_{c}$ is an unbiased estimator of $V_{c}$, with a
binomial standard error
\begin{equation}
  \sigma_{c}=\sqrt{\frac{f_{c}(1-f_{c})}{N_{\mathrm{ind}}}}\le\frac{1}{2\sqrt{N_{\mathrm{ind}}}},
  \label{eq:binomial}
\end{equation}
where $N_{\mathrm{ind}}=N/2=2048$ is the number of independently
integrated initial conditions; stratification makes
Eq.~\eqref{eq:binomial} a conservative bound. Monte Carlo estimates of
basin volumes remain meaningful when the basin boundaries are fractal,
even though the fate of individual initial conditions near them is not
resolved, and reach their limits only for riddled
basins~\cite{schultz2017potentials}, which require an invariant
subspace and do not arise here. In the figures each bar
corresponds to a single value of the control parameter, so that
$\sigma_{c}\le1.1\times10^{-2}$.

The two cuts are sampled on $1001$ values of $\gamma\in[0.20,1.00]$
($\Delta\gamma=8\times10^{-4}$) at fixed $\omega=1.3$ and on $791$ values
of $\omega\in[0.05,4.00]$ ($\Delta\omega=5\times10^{-3}$) at fixed
$\gamma=1$, i.e.\ $2048$ trajectories of duration
$T_{\mathrm{tr}}+T_{\mathrm{int}}+T_{\lambda}$ per parameter value.
The unclassified fraction,
which collects the trajectories whose transient outlasts the
classification window, is reported explicitly rather than redistributed
among the other classes; it is a useful diagnostic, since it peaks where
chaotic transients are long, typically at window boundaries.

\subsection{Continuation sweeps}
\label{subsec:continuation}

The bifurcation diagrams of Figs.~\ref{fig:fractions_gamma}(b)
and~\ref{fig:fractions_omega}(b) are computed on the same parameter
grids as the samples, with the same classifier and the same integration
windows. A single trajectory is started at the lowest value of the
control parameter from the reference initial condition
$(x(0),\dot{x}(0),\phi_{0})=(2.0,0.1,0)$ and followed upward: at each
subsequent value the integration is seeded with the final state reached
at the previous one, handed over coherently with the forcing, every
value being integrated over an integer number of driving periods so that
the inherited state is always at the same drive phase.

The same procedure, applied column by column, produces the map of the
parameter plane of Sec.~\ref{sec:plane}: at each of the $800$
frequencies of the grid the continuation runs upward in $\gamma$ over
$800$ values, seeded at the lowest amplitude from the reference initial
condition, so that the seed is reset at every column. For this map the
wells visited are determined over the final segment of the recorded
window, of length $T_{\mathrm{w}}=100$ or, if longer, one orbital period
$kT_{d}$ for periodic orbits and ten drive periods for chaotic ones, so
that the segment covers at least one full excursion of the orbit also at
the lowest frequencies.

\bibliography{/home/arturo/Dropbox/bibtex/mybib}

\begin{thebibliography}{42}%
\makeatletter
\providecommand \@ifxundefined [1]{%
 \@ifx{#1\undefined}
}%
\providecommand \@ifnum [1]{%
 \ifnum #1\expandafter \@firstoftwo
 \else \expandafter \@secondoftwo
 \fi
}%
\providecommand \@ifx [1]{%
 \ifx #1\expandafter \@firstoftwo
 \else \expandafter \@secondoftwo
 \fi
}%
\providecommand \natexlab [1]{#1}%
\providecommand \enquote  [1]{``#1''}%
\providecommand \bibnamefont  [1]{#1}%
\providecommand \bibfnamefont [1]{#1}%
\providecommand \citenamefont [1]{#1}%
\providecommand \href@noop [0]{\@secondoftwo}%
\providecommand \href [0]{\begingroup \@sanitize@url \@href}%
\providecommand \@href[1]{\@@startlink{#1}\@@href}%
\providecommand \@@href[1]{\endgroup#1\@@endlink}%
\providecommand \@sanitize@url [0]{\catcode `\\12\catcode `\$12\catcode
  `\&12\catcode `\#12\catcode `\^12\catcode `\_12\catcode `\%12\relax}%
\providecommand \@@startlink[1]{}%
\providecommand \@@endlink[0]{}%
\providecommand \url  [0]{\begingroup\@sanitize@url \@url }%
\providecommand \@url [1]{\endgroup\@href {#1}{\urlprefix }}%
\providecommand \urlprefix  [0]{URL }%
\providecommand \Eprint [0]{\href }%
\providecommand \doibase [0]{https://doi.org/}%
\providecommand \selectlanguage [0]{\@gobble}%
\providecommand \bibinfo  [0]{\@secondoftwo}%
\providecommand \bibfield  [0]{\@secondoftwo}%
\providecommand \translation [1]{[#1]}%
\providecommand \BibitemOpen [0]{}%
\providecommand \bibitemStop [0]{}%
\providecommand \bibitemNoStop [0]{.\EOS\space}%
\providecommand \EOS [0]{\spacefactor3000\relax}%
\providecommand \BibitemShut  [1]{\csname bibitem#1\endcsname}%
\let\auto@bib@innerbib\@empty
\bibitem [{\citenamefont {Feudel}(2008)}]{feudel2008complex}%
  \BibitemOpen
  \bibfield  {author} {\bibinfo {author} {\bibfnamefont {U.}~\bibnamefont
  {Feudel}},\ }\bibfield  {title} {\enquote {\bibinfo {title} {Complex dynamics
  in multistable systems},}\ }\href {https://doi.org/10.1142/S0218127408021233}
  {\bibfield  {journal} {\bibinfo  {journal} {International Journal of
  Bifurcation and Chaos}\ }\textbf {\bibinfo {volume} {18}},\ \bibinfo {pages}
  {1607--1626} (\bibinfo {year} {2008})}\BibitemShut {NoStop}%
\bibitem [{\citenamefont {Pisarchik}\ and\ \citenamefont
  {Feudel}(2014)}]{pisarchik2014control}%
  \BibitemOpen
  \bibfield  {author} {\bibinfo {author} {\bibfnamefont {A.~N.}\ \bibnamefont
  {Pisarchik}}\ and\ \bibinfo {author} {\bibfnamefont {U.}~\bibnamefont
  {Feudel}},\ }\bibfield  {title} {\enquote {\bibinfo {title} {Control of
  multistability},}\ }\href {https://doi.org/10.1016/j.physrep.2014.02.007}
  {\bibfield  {journal} {\bibinfo  {journal} {Physics Reports}\ }\textbf
  {\bibinfo {volume} {540}},\ \bibinfo {pages} {167--218} (\bibinfo {year}
  {2014})}\BibitemShut {NoStop}%
\bibitem [{\citenamefont {Grebogi}\ \emph {et~al.}(1983)\citenamefont
  {Grebogi}, \citenamefont {McDonald}, \citenamefont {Ott},\ and\ \citenamefont
  {Yorke}}]{grebogi1983final}%
  \BibitemOpen
  \bibfield  {author} {\bibinfo {author} {\bibfnamefont {C.}~\bibnamefont
  {Grebogi}}, \bibinfo {author} {\bibfnamefont {S.~W.}\ \bibnamefont
  {McDonald}}, \bibinfo {author} {\bibfnamefont {E.}~\bibnamefont {Ott}},\ and\
  \bibinfo {author} {\bibfnamefont {J.~A.}\ \bibnamefont {Yorke}},\ }\bibfield
  {title} {\enquote {\bibinfo {title} {Final state sensitivity: An obstruction
  to predictability},}\ }\href {https://doi.org/10.1016/0375-9601(83)90945-3}
  {\bibfield  {journal} {\bibinfo  {journal} {Physics Letters A}\ }\textbf
  {\bibinfo {volume} {99}},\ \bibinfo {pages} {415--418} (\bibinfo {year}
  {1983})}\BibitemShut {NoStop}%
\bibitem [{\citenamefont {McDonald}\ \emph {et~al.}(1985)\citenamefont
  {McDonald}, \citenamefont {Grebogi}, \citenamefont {Ott},\ and\ \citenamefont
  {Yorke}}]{mcdonald1985fractal}%
  \BibitemOpen
  \bibfield  {author} {\bibinfo {author} {\bibfnamefont {S.~W.}\ \bibnamefont
  {McDonald}}, \bibinfo {author} {\bibfnamefont {C.}~\bibnamefont {Grebogi}},
  \bibinfo {author} {\bibfnamefont {E.}~\bibnamefont {Ott}},\ and\ \bibinfo
  {author} {\bibfnamefont {J.~A.}\ \bibnamefont {Yorke}},\ }\bibfield  {title}
  {\enquote {\bibinfo {title} {Fractal basin boundaries},}\ }\href
  {https://doi.org/10.1016/0167-2789(85)90001-6} {\bibfield  {journal}
  {\bibinfo  {journal} {Physica D}\ }\textbf {\bibinfo {volume} {17}},\
  \bibinfo {pages} {125--153} (\bibinfo {year} {1985})}\BibitemShut {NoStop}%
\bibitem [{\citenamefont {Aguirre}, \citenamefont {Viana},\ and\ \citenamefont
  {Sanju\'an}(2009)}]{aguirre2009fractal}%
  \BibitemOpen
  \bibfield  {author} {\bibinfo {author} {\bibfnamefont {J.}~\bibnamefont
  {Aguirre}}, \bibinfo {author} {\bibfnamefont {R.~L.}\ \bibnamefont {Viana}},\
  and\ \bibinfo {author} {\bibfnamefont {M.~A.~F.}\ \bibnamefont {Sanju\'an}},\
  }\bibfield  {title} {\enquote {\bibinfo {title} {Fractal structures in
  nonlinear dynamics},}\ }\href {https://doi.org/10.1103/RevModPhys.81.333}
  {\bibfield  {journal} {\bibinfo  {journal} {Reviews of Modern Physics}\
  }\textbf {\bibinfo {volume} {81}},\ \bibinfo {pages} {333--386} (\bibinfo
  {year} {2009})}\BibitemShut {NoStop}%
\bibitem [{\citenamefont {Daza}\ \emph {et~al.}(2016)\citenamefont {Daza},
  \citenamefont {Wagemakers}, \citenamefont {Georgeot}, \citenamefont
  {Gu{\'e}ry-Odelin},\ and\ \citenamefont {Sanju{\'a}n}}]{daza2016basin}%
  \BibitemOpen
  \bibfield  {author} {\bibinfo {author} {\bibfnamefont {A.}~\bibnamefont
  {Daza}}, \bibinfo {author} {\bibfnamefont {A.}~\bibnamefont {Wagemakers}},
  \bibinfo {author} {\bibfnamefont {B.}~\bibnamefont {Georgeot}}, \bibinfo
  {author} {\bibfnamefont {D.}~\bibnamefont {Gu{\'e}ry-Odelin}},\ and\ \bibinfo
  {author} {\bibfnamefont {M.~A.~F.}\ \bibnamefont {Sanju{\'a}n}},\ }\bibfield
  {title} {\enquote {\bibinfo {title} {Basin entropy: a new tool to analyze
  uncertainty in dynamical systems},}\ }\href@noop {} {\bibfield  {journal}
  {\bibinfo  {journal} {Sci. Rep.}\ }\textbf {\bibinfo {volume} {6}},\ \bibinfo
  {pages} {1--10} (\bibinfo {year} {2016})}\BibitemShut {NoStop}%
\bibitem [{\citenamefont {Daza}, \citenamefont {Wagemakers},\ and\
  \citenamefont {Sanju{\'a}n}(2023)}]{daza2023unpredictability}%
  \BibitemOpen
  \bibfield  {author} {\bibinfo {author} {\bibfnamefont {A.}~\bibnamefont
  {Daza}}, \bibinfo {author} {\bibfnamefont {A.}~\bibnamefont {Wagemakers}},\
  and\ \bibinfo {author} {\bibfnamefont {M.~A.~F.}\ \bibnamefont
  {Sanju{\'a}n}},\ }\bibfield  {title} {\enquote {\bibinfo {title}
  {Unpredictability and basin entropy},}\ }\href@noop {} {\bibfield  {journal}
  {\bibinfo  {journal} {Europhysics Letters}\ }\textbf {\bibinfo {volume}
  {141}},\ \bibinfo {pages} {43001} (\bibinfo {year} {2023})}\BibitemShut
  {NoStop}%
\bibitem [{\citenamefont {Menck}\ \emph {et~al.}(2013)\citenamefont {Menck},
  \citenamefont {Heitzig}, \citenamefont {Marwan},\ and\ \citenamefont
  {Kurths}}]{menck2013basin}%
  \BibitemOpen
  \bibfield  {author} {\bibinfo {author} {\bibfnamefont {P.~J.}\ \bibnamefont
  {Menck}}, \bibinfo {author} {\bibfnamefont {J.}~\bibnamefont {Heitzig}},
  \bibinfo {author} {\bibfnamefont {N.}~\bibnamefont {Marwan}},\ and\ \bibinfo
  {author} {\bibfnamefont {J.}~\bibnamefont {Kurths}},\ }\bibfield  {title}
  {\enquote {\bibinfo {title} {How basin stability complements the
  linear-stability paradigm},}\ }\href {https://doi.org/10.1038/nphys2516}
  {\bibfield  {journal} {\bibinfo  {journal} {Nature Physics}\ }\textbf
  {\bibinfo {volume} {9}},\ \bibinfo {pages} {89--92} (\bibinfo {year}
  {2013})}\BibitemShut {NoStop}%
\bibitem [{\citenamefont {Gelbrecht}, \citenamefont {Kurths},\ and\
  \citenamefont {Hellmann}(2020)}]{gelbrecht2020monte}%
  \BibitemOpen
  \bibfield  {author} {\bibinfo {author} {\bibfnamefont {M.}~\bibnamefont
  {Gelbrecht}}, \bibinfo {author} {\bibfnamefont {J.}~\bibnamefont {Kurths}},\
  and\ \bibinfo {author} {\bibfnamefont {F.}~\bibnamefont {Hellmann}},\
  }\bibfield  {title} {\enquote {\bibinfo {title} {Monte {C}arlo basin
  bifurcation analysis},}\ }\href {https://doi.org/10.1088/1367-2630/ab7a05}
  {\bibfield  {journal} {\bibinfo  {journal} {New Journal of Physics}\ }\textbf
  {\bibinfo {volume} {22}},\ \bibinfo {pages} {033032} (\bibinfo {year}
  {2020})}\BibitemShut {NoStop}%
\bibitem [{\citenamefont {Stender}\ and\ \citenamefont
  {Hoffmann}(2022)}]{stender2022bstab}%
  \BibitemOpen
  \bibfield  {author} {\bibinfo {author} {\bibfnamefont {M.}~\bibnamefont
  {Stender}}\ and\ \bibinfo {author} {\bibfnamefont {N.}~\bibnamefont
  {Hoffmann}},\ }\bibfield  {title} {\enquote {\bibinfo {title} {{bSTAB}: an
  open-source software for computing the basin stability of multi-stable
  dynamical systems},}\ }\href {https://doi.org/10.1007/s11071-021-06786-5}
  {\bibfield  {journal} {\bibinfo  {journal} {Nonlinear Dynamics}\ }\textbf
  {\bibinfo {volume} {107}},\ \bibinfo {pages} {1451--1468} (\bibinfo {year}
  {2022})}\BibitemShut {NoStop}%
\bibitem [{\citenamefont {Datseris}, \citenamefont {Rossi},\ and\ \citenamefont
  {Wagemakers}(2023)}]{datseris2023framework}%
  \BibitemOpen
  \bibfield  {author} {\bibinfo {author} {\bibfnamefont {G.}~\bibnamefont
  {Datseris}}, \bibinfo {author} {\bibfnamefont {K.~L.}\ \bibnamefont
  {Rossi}},\ and\ \bibinfo {author} {\bibfnamefont {A.}~\bibnamefont
  {Wagemakers}},\ }\bibfield  {title} {\enquote {\bibinfo {title} {Framework
  for global stability analysis of dynamical systems},}\ }\href
  {https://doi.org/10.1063/5.0159675} {\bibfield  {journal} {\bibinfo
  {journal} {Chaos: An Interdisciplinary Journal of Nonlinear Science}\
  }\textbf {\bibinfo {volume} {33}},\ \bibinfo {pages} {073151} (\bibinfo
  {year} {2023})}\BibitemShut {NoStop}%
\bibitem [{\citenamefont {Wagemakers}(2025)}]{wagemakers2025basins}%
  \BibitemOpen
  \bibfield  {author} {\bibinfo {author} {\bibfnamefont {A.}~\bibnamefont
  {Wagemakers}},\ }\bibfield  {title} {\enquote {\bibinfo {title} {Basins of
  attraction: A dynamical zoo},}\ }\href@noop {} {\bibfield  {journal}
  {\bibinfo  {journal} {International Journal of Bifurcation and Chaos}\
  }\textbf {\bibinfo {volume} {35}},\ \bibinfo {pages} {2530024} (\bibinfo
  {year} {2025})}\BibitemShut {NoStop}%
\bibitem [{\citenamefont {Holmes}(1979)}]{holmes1979nonlinear}%
  \BibitemOpen
  \bibfield  {author} {\bibinfo {author} {\bibfnamefont {P.~J.}\ \bibnamefont
  {Holmes}},\ }\bibfield  {title} {\enquote {\bibinfo {title} {A nonlinear
  oscillator with a strange attractor},}\ }\href@noop {} {\bibfield  {journal}
  {\bibinfo  {journal} {Philosophical Transactions of the Royal Society A}\
  }\textbf {\bibinfo {volume} {292}},\ \bibinfo {pages} {419--448} (\bibinfo
  {year} {1979})}\BibitemShut {NoStop}%
\bibitem [{\citenamefont {Ueda}(1980)}]{ueda1980}%
  \BibitemOpen
  \bibfield  {author} {\bibinfo {author} {\bibfnamefont {Y.}~\bibnamefont
  {Ueda}},\ }\bibfield  {title} {\enquote {\bibinfo {title} {Steady motions
  exhibited by {D}uffing's equation},}\ }\href@noop {} {\bibfield  {journal}
  {\bibinfo  {journal} {Journal of the Physical Society of Japan}\ }\textbf
  {\bibinfo {volume} {30}},\ \bibinfo {pages} {440--446} (\bibinfo {year}
  {1980})}\BibitemShut {NoStop}%
\bibitem [{\citenamefont {Korsch}, \citenamefont {Jodl},\ and\ \citenamefont
  {Hartmann}(2008)}]{ChaosKorsch}%
  \BibitemOpen
  \bibfield  {author} {\bibinfo {author} {\bibfnamefont {H.~J.}\ \bibnamefont
  {Korsch}}, \bibinfo {author} {\bibfnamefont {H.-J.}\ \bibnamefont {Jodl}},\
  and\ \bibinfo {author} {\bibfnamefont {T.}~\bibnamefont {Hartmann}},\
  }\href@noop {} {\emph {\bibinfo {title} {Chaos: A Program Collection for the
  {PC}}}},\ \bibinfo {edition} {3rd}\ ed.\ (\bibinfo  {publisher}
  {Springer-Verlag},\ \bibinfo {address} {Berlin Heidelberg},\ \bibinfo {year}
  {2008})\BibitemShut {NoStop}%
\bibitem [{\citenamefont {Strogatz}(2018)}]{strogatz2018nonlinear}%
  \BibitemOpen
  \bibfield  {author} {\bibinfo {author} {\bibfnamefont {S.~H.}\ \bibnamefont
  {Strogatz}},\ }\href@noop {} {\emph {\bibinfo {title} {Nonlinear dynamics and
  chaos with student solutions manual: With applications to physics, biology,
  chemistry, and engineering}}}\ (\bibinfo  {publisher} {CRC press},\ \bibinfo
  {year} {2018})\BibitemShut {NoStop}%
\bibitem [{\citenamefont {Argyris}\ \emph {et~al.}(2015)\citenamefont
  {Argyris}, \citenamefont {Faust}, \citenamefont {Haase},\ and\ \citenamefont
  {Friedrich}}]{argyris2015exploration}%
  \BibitemOpen
  \bibfield  {author} {\bibinfo {author} {\bibfnamefont {J.~H.}\ \bibnamefont
  {Argyris}}, \bibinfo {author} {\bibfnamefont {G.}~\bibnamefont {Faust}},
  \bibinfo {author} {\bibfnamefont {M.}~\bibnamefont {Haase}},\ and\ \bibinfo
  {author} {\bibfnamefont {R.}~\bibnamefont {Friedrich}},\ }\href@noop {}
  {\emph {\bibinfo {title} {An exploration of dynamical systems and chaos:
  completely revised and enlarged second edition}}}\ (\bibinfo  {publisher}
  {Springer},\ \bibinfo {year} {2015})\BibitemShut {NoStop}%
\bibitem [{\citenamefont {Moon}\ and\ \citenamefont
  {Li}(1985)}]{moon1985fractal}%
  \BibitemOpen
  \bibfield  {author} {\bibinfo {author} {\bibfnamefont {F.~C.}\ \bibnamefont
  {Moon}}\ and\ \bibinfo {author} {\bibfnamefont {G.-X.}\ \bibnamefont {Li}},\
  }\bibfield  {title} {\enquote {\bibinfo {title} {Fractal basin boundaries and
  homoclinic orbits for periodic motion in a two-well potential},}\ }\href
  {https://doi.org/10.1103/PhysRevLett.55.1439} {\bibfield  {journal} {\bibinfo
   {journal} {Physical Review Letters}\ }\textbf {\bibinfo {volume} {55}},\
  \bibinfo {pages} {1439--1442} (\bibinfo {year} {1985})}\BibitemShut {NoStop}%
\bibitem [{\citenamefont {Ueda}\ \emph {et~al.}(1990)\citenamefont {Ueda},
  \citenamefont {Yoshida}, \citenamefont {Stewart},\ and\ \citenamefont
  {Thompson}}]{ueda1990basin}%
  \BibitemOpen
  \bibfield  {author} {\bibinfo {author} {\bibfnamefont {Y.}~\bibnamefont
  {Ueda}}, \bibinfo {author} {\bibfnamefont {S.}~\bibnamefont {Yoshida}},
  \bibinfo {author} {\bibfnamefont {H.~B.}\ \bibnamefont {Stewart}},\ and\
  \bibinfo {author} {\bibfnamefont {J.~M.~T.}\ \bibnamefont {Thompson}},\
  }\bibfield  {title} {\enquote {\bibinfo {title} {Basin explosions and escape
  phenomena in the twin-well {D}uffing oscillator: compound global bifurcations
  organizing behaviour},}\ }\href {https://doi.org/10.1098/rsta.1990.0107}
  {\bibfield  {journal} {\bibinfo  {journal} {Philosophical Transactions of the
  Royal Society of London A}\ }\textbf {\bibinfo {volume} {332}},\ \bibinfo
  {pages} {169--186} (\bibinfo {year} {1990})}\BibitemShut {NoStop}%
\bibitem [{\citenamefont {Aguirre}\ and\ \citenamefont
  {Sanju\'an}(2002)}]{aguirre2002unpredictable}%
  \BibitemOpen
  \bibfield  {author} {\bibinfo {author} {\bibfnamefont {J.}~\bibnamefont
  {Aguirre}}\ and\ \bibinfo {author} {\bibfnamefont {M.~A.~F.}\ \bibnamefont
  {Sanju\'an}},\ }\bibfield  {title} {\enquote {\bibinfo {title} {Unpredictable
  behavior in the {D}uffing oscillator: {W}ada basins},}\ }\href
  {https://doi.org/10.1016/S0167-2789(02)00565-1} {\bibfield  {journal}
  {\bibinfo  {journal} {Physica D}\ }\textbf {\bibinfo {volume} {171}},\
  \bibinfo {pages} {41--51} (\bibinfo {year} {2002})}\BibitemShut {NoStop}%
\bibitem [{\citenamefont {Bonatto}\ and\ \citenamefont
  {Gallas}(2008)}]{bonatto2008chaotic}%
  \BibitemOpen
  \bibfield  {author} {\bibinfo {author} {\bibfnamefont {C.}~\bibnamefont
  {Bonatto}}\ and\ \bibinfo {author} {\bibfnamefont {J.~A.~C.}\ \bibnamefont
  {Gallas}},\ }\bibfield  {title} {\enquote {\bibinfo {title} {Chaotic phase
  synchronization},}\ }\href@noop {} {\bibfield  {journal} {\bibinfo  {journal}
  {Physical Review E}\ }\textbf {\bibinfo {volume} {75}},\ \bibinfo {pages}
  {055217} (\bibinfo {year} {2008})}\BibitemShut {NoStop}%
\bibitem [{\citenamefont {Hasan}\ \emph {et~al.}(2023)\citenamefont {Hasan},
  \citenamefont {Greenwood}, \citenamefont {Parker}, \citenamefont {Kong},\
  and\ \citenamefont {Wang}}]{hasan2023fractal}%
  \BibitemOpen
  \bibfield  {author} {\bibinfo {author} {\bibfnamefont {M.~N.}\ \bibnamefont
  {Hasan}}, \bibinfo {author} {\bibfnamefont {T.~E.}\ \bibnamefont
  {Greenwood}}, \bibinfo {author} {\bibfnamefont {R.~G.}\ \bibnamefont
  {Parker}}, \bibinfo {author} {\bibfnamefont {Y.~L.}\ \bibnamefont {Kong}},\
  and\ \bibinfo {author} {\bibfnamefont {P.}~\bibnamefont {Wang}},\ }\bibfield
  {title} {\enquote {\bibinfo {title} {Fractal patterns in the parameter space
  of a bistable {D}uffing oscillator},}\ }\href
  {https://doi.org/10.1103/PhysRevE.108.L022201} {\bibfield  {journal}
  {\bibinfo  {journal} {Physical Review E}\ }\textbf {\bibinfo {volume}
  {108}},\ \bibinfo {pages} {L022201} (\bibinfo {year} {2023})}\BibitemShut
  {NoStop}%
\bibitem [{\citenamefont {Todd}\ and\ \citenamefont
  {Virgin}(1997)}]{todd1997experimental}%
  \BibitemOpen
  \bibfield  {author} {\bibinfo {author} {\bibfnamefont {M.~D.}\ \bibnamefont
  {Todd}}\ and\ \bibinfo {author} {\bibfnamefont {L.~N.}\ \bibnamefont
  {Virgin}},\ }\bibfield  {title} {\enquote {\bibinfo {title} {An experimental
  verification of basin metamorphoses in a nonlinear mechanical system},}\
  }\href {https://doi.org/10.1142/S0218127497001060} {\bibfield  {journal}
  {\bibinfo  {journal} {International Journal of Bifurcation and Chaos}\
  }\textbf {\bibinfo {volume} {7}},\ \bibinfo {pages} {1337--1357} (\bibinfo
  {year} {1997})}\BibitemShut {NoStop}%
\bibitem [{\citenamefont {Gargiulo}\ \emph {et~al.}(2026)\citenamefont
  {Gargiulo}, \citenamefont {Gutiérrez}, \citenamefont {Tarigo}, \citenamefont
  {Stari},\ and\ \citenamefont {Martí}}]{gargiulo2026order}%
  \BibitemOpen
  \bibfield  {author} {\bibinfo {author} {\bibfnamefont {P.~R.}\ \bibnamefont
  {Gargiulo}}, \bibinfo {author} {\bibfnamefont {C.}~\bibnamefont
  {Gutiérrez}}, \bibinfo {author} {\bibfnamefont {J.~P.}\ \bibnamefont
  {Tarigo}}, \bibinfo {author} {\bibfnamefont {C.}~\bibnamefont {Stari}},\ and\
  \bibinfo {author} {\bibfnamefont {A.~C.}\ \bibnamefont {Martí}},\ }\bibfield
   {title} {\enquote {\bibinfo {title} {Experimental observation of chaotic and
  multistable dynamics in a duffing–holmes-type analog circuit:
  Antiperiodicity and attractor-coexistence signatures},}\ }\href
  {https://doi.org/https://doi.org/10.1016/j.chaos.2026.118992} {\bibfield
  {journal} {\bibinfo  {journal} {Chaos, Solitons \& Fractals}\ }\textbf
  {\bibinfo {volume} {212}},\ \bibinfo {pages} {118992} (\bibinfo {year}
  {2026})}\BibitemShut {NoStop}%
\bibitem [{\citenamefont {Mart\'i}\ and\ \citenamefont
  {Leonel}(2026)}]{marti2026antiperiodicity}%
  \BibitemOpen
  \bibfield  {author} {\bibinfo {author} {\bibfnamefont {A.~C.}\ \bibnamefont
  {Mart\'i}}\ and\ \bibinfo {author} {\bibfnamefont {E.~D.}\ \bibnamefont
  {Leonel}},\ }\bibfield  {title} {\enquote {\bibinfo {title} {Antiperiodicity
  in the {D}uffing--{H}olmes oscillator: symmetry origin, parity selection, and
  transition to broken symmetry},}\ }\href@noop {} {\bibfield  {journal}
  {\bibinfo  {journal} {submitted}\ } (\bibinfo {year} {2026})},\ \bibinfo
  {note} {arXiv:2607.05604}\BibitemShut {NoStop}%
\bibitem [{\citenamefont {Swift}\ and\ \citenamefont
  {Wiesenfeld}(1984)}]{swift1984suppression}%
  \BibitemOpen
  \bibfield  {author} {\bibinfo {author} {\bibfnamefont {J.~W.}\ \bibnamefont
  {Swift}}\ and\ \bibinfo {author} {\bibfnamefont {K.}~\bibnamefont
  {Wiesenfeld}},\ }\bibfield  {title} {\enquote {\bibinfo {title} {Suppression
  of period doubling in symmetric systems},}\ }\href
  {https://doi.org/10.1103/PhysRevLett.52.705} {\bibfield  {journal} {\bibinfo
  {journal} {Phys. Rev. Lett.}\ }\textbf {\bibinfo {volume} {52}},\ \bibinfo
  {pages} {705} (\bibinfo {year} {1984})}\BibitemShut {NoStop}%
\bibitem [{\citenamefont {Chossat}\ and\ \citenamefont
  {Golubitsky}(1988)}]{chossat1988symmetry}%
  \BibitemOpen
  \bibfield  {author} {\bibinfo {author} {\bibfnamefont {P.}~\bibnamefont
  {Chossat}}\ and\ \bibinfo {author} {\bibfnamefont {M.}~\bibnamefont
  {Golubitsky}},\ }\bibfield  {title} {\enquote {\bibinfo {title}
  {Symmetry-increasing bifurcation of chaotic attractors},}\ }\href
  {https://doi.org/10.1016/0167-2789(88)90066-8} {\bibfield  {journal}
  {\bibinfo  {journal} {Physica D}\ }\textbf {\bibinfo {volume} {32}},\
  \bibinfo {pages} {423--436} (\bibinfo {year} {1988})}\BibitemShut {NoStop}%
\bibitem [{\citenamefont {Melbourne}, \citenamefont {Dellnitz},\ and\
  \citenamefont {Golubitsky}(1993)}]{melbourne1993structure}%
  \BibitemOpen
  \bibfield  {author} {\bibinfo {author} {\bibfnamefont {I.}~\bibnamefont
  {Melbourne}}, \bibinfo {author} {\bibfnamefont {M.}~\bibnamefont
  {Dellnitz}},\ and\ \bibinfo {author} {\bibfnamefont {M.}~\bibnamefont
  {Golubitsky}},\ }\bibfield  {title} {\enquote {\bibinfo {title} {The
  structure of symmetric attractors},}\ }\href
  {https://doi.org/10.1007/BF00386369} {\bibfield  {journal} {\bibinfo
  {journal} {Arch. Ration. Mech. Anal.}\ }\textbf {\bibinfo {volume} {123}},\
  \bibinfo {pages} {75--98} (\bibinfo {year} {1993})}\BibitemShut {NoStop}%
\bibitem [{\citenamefont {Xie}(2025)}]{xie2025symmetry}%
  \BibitemOpen
  \bibfield  {author} {\bibinfo {author} {\bibfnamefont {X.}~\bibnamefont
  {Xie}},\ }\href@noop {} {\enquote {\bibinfo {title} {Symmetry groups of
  basins of attraction in equivariant dynamical systems},}\ } (\bibinfo {year}
  {2025}),\ \bibinfo {note} {arXiv:2509.11403},\ \Eprint
  {https://arxiv.org/abs/2509.11403} {arXiv:2509.11403 [nlin.CD]} \BibitemShut
  {NoStop}%
\bibitem [{\citenamefont {Tarigo}\ \emph {et~al.}(2024)\citenamefont {Tarigo},
  \citenamefont {Stari}, \citenamefont {Masoller},\ and\ \citenamefont
  {Martí}}]{tarigo2024basin}%
  \BibitemOpen
  \bibfield  {author} {\bibinfo {author} {\bibfnamefont {J.~P.}\ \bibnamefont
  {Tarigo}}, \bibinfo {author} {\bibfnamefont {C.}~\bibnamefont {Stari}},
  \bibinfo {author} {\bibfnamefont {C.}~\bibnamefont {Masoller}},\ and\
  \bibinfo {author} {\bibfnamefont {A.~C.}\ \bibnamefont {Martí}},\ }\bibfield
   {title} {\enquote {\bibinfo {title} {Basin entropy as an indicator of a
  bifurcation in a time-delayed system},}\ }\href
  {https://doi.org/10.1063/5.0201932} {\bibfield  {journal} {\bibinfo
  {journal} {Chaos: An Interdisciplinary Journal of Nonlinear Science}\
  }\textbf {\bibinfo {volume} {34}},\ \bibinfo {pages} {053113} (\bibinfo
  {year} {2024})}\BibitemShut {NoStop}%
\bibitem [{\citenamefont {Tarigo}, \citenamefont {Stari},\ and\ \citenamefont
  {Martí}(2024)}]{tarigo2024basinMG}%
  \BibitemOpen
  \bibfield  {author} {\bibinfo {author} {\bibfnamefont {J.~P.}\ \bibnamefont
  {Tarigo}}, \bibinfo {author} {\bibfnamefont {C.}~\bibnamefont {Stari}},\ and\
  \bibinfo {author} {\bibfnamefont {A.~C.}\ \bibnamefont {Martí}},\ }\bibfield
   {title} {\enquote {\bibinfo {title} {Basin of attraction organization in
  infinite-dimensional delayed systems: A stochastic basin entropy approach},}\
  }\href {https://doi.org/10.1063/5.0234028} {\bibfield  {journal} {\bibinfo
  {journal} {Chaos: An Interdisciplinary Journal of Nonlinear Science}\
  }\textbf {\bibinfo {volume} {34}},\ \bibinfo {pages} {123125} (\bibinfo
  {year} {2024})}\BibitemShut {NoStop}%
\bibitem [{\citenamefont {Soliman}\ and\ \citenamefont
  {Thompson}(1989)}]{soliman1989integrity}%
  \BibitemOpen
  \bibfield  {author} {\bibinfo {author} {\bibfnamefont {M.~S.}\ \bibnamefont
  {Soliman}}\ and\ \bibinfo {author} {\bibfnamefont {J.~M.~T.}\ \bibnamefont
  {Thompson}},\ }\bibfield  {title} {\enquote {\bibinfo {title} {Integrity
  measures quantifying the erosion of smooth and fractal basins of
  attraction},}\ }\href {https://doi.org/10.1016/0022-460X(89)90699-8}
  {\bibfield  {journal} {\bibinfo  {journal} {Journal of Sound and Vibration}\
  }\textbf {\bibinfo {volume} {135}},\ \bibinfo {pages} {453--475} (\bibinfo
  {year} {1989})}\BibitemShut {NoStop}%
\bibitem [{\citenamefont {Rega}\ and\ \citenamefont
  {Lenci}(2008)}]{rega2008dynamical}%
  \BibitemOpen
  \bibfield  {author} {\bibinfo {author} {\bibfnamefont {G.}~\bibnamefont
  {Rega}}\ and\ \bibinfo {author} {\bibfnamefont {S.}~\bibnamefont {Lenci}},\
  }\bibfield  {title} {\enquote {\bibinfo {title} {Dynamical integrity and
  control of nonlinear mechanical oscillators},}\ }\href
  {https://doi.org/10.1177/1077546307079403} {\bibfield  {journal} {\bibinfo
  {journal} {Journal of Vibration and Control}\ }\textbf {\bibinfo {volume}
  {14}},\ \bibinfo {pages} {159--179} (\bibinfo {year} {2008})}\BibitemShut
  {NoStop}%
\bibitem [{\citenamefont {Ishii}, \citenamefont {Fujisaka},\ and\ \citenamefont
  {Inoue}(1986)}]{ishii1986breakdown}%
  \BibitemOpen
  \bibfield  {author} {\bibinfo {author} {\bibfnamefont {H.}~\bibnamefont
  {Ishii}}, \bibinfo {author} {\bibfnamefont {H.}~\bibnamefont {Fujisaka}},\
  and\ \bibinfo {author} {\bibfnamefont {M.}~\bibnamefont {Inoue}},\ }\bibfield
   {title} {\enquote {\bibinfo {title} {Breakdown of chaos symmetry and
  intermittency in the double-well potential system},}\ }\href@noop {}
  {\bibfield  {journal} {\bibinfo  {journal} {Phys. Lett. A}\ }\textbf
  {\bibinfo {volume} {116}},\ \bibinfo {pages} {257} (\bibinfo {year}
  {1986})}\BibitemShut {NoStop}%
\bibitem [{\citenamefont {Grebogi}\ \emph {et~al.}(1987)\citenamefont
  {Grebogi}, \citenamefont {Ott}, \citenamefont {Romeiras},\ and\ \citenamefont
  {Yorke}}]{grebogi1987critical}%
  \BibitemOpen
  \bibfield  {author} {\bibinfo {author} {\bibfnamefont {C.}~\bibnamefont
  {Grebogi}}, \bibinfo {author} {\bibfnamefont {E.}~\bibnamefont {Ott}},
  \bibinfo {author} {\bibfnamefont {F.}~\bibnamefont {Romeiras}},\ and\
  \bibinfo {author} {\bibfnamefont {J.~A.}\ \bibnamefont {Yorke}},\ }\bibfield
  {title} {\enquote {\bibinfo {title} {Critical exponents for crisis-induced
  intermittency},}\ }\href {https://doi.org/10.1103/PhysRevA.36.5365}
  {\bibfield  {journal} {\bibinfo  {journal} {Phys. Rev. A}\ }\textbf {\bibinfo
  {volume} {36}},\ \bibinfo {pages} {5365--5380} (\bibinfo {year}
  {1987})}\BibitemShut {NoStop}%
\bibitem [{\citenamefont {Puy}\ \emph {et~al.}(2021)\citenamefont {Puy},
  \citenamefont {Daza}, \citenamefont {Wagemakers},\ and\ \citenamefont
  {Sanju\'an}}]{puy2021test}%
  \BibitemOpen
  \bibfield  {author} {\bibinfo {author} {\bibfnamefont {A.}~\bibnamefont
  {Puy}}, \bibinfo {author} {\bibfnamefont {A.}~\bibnamefont {Daza}}, \bibinfo
  {author} {\bibfnamefont {A.}~\bibnamefont {Wagemakers}},\ and\ \bibinfo
  {author} {\bibfnamefont {M.~A.~F.}\ \bibnamefont {Sanju\'an}},\ }\bibfield
  {title} {\enquote {\bibinfo {title} {A test for fractal boundaries based on
  the basin entropy},}\ }\href {https://doi.org/10.1016/j.cnsns.2020.105588}
  {\bibfield  {journal} {\bibinfo  {journal} {Communications in Nonlinear
  Science and Numerical Simulation}\ }\textbf {\bibinfo {volume} {95}},\
  \bibinfo {pages} {105588} (\bibinfo {year} {2021})}\BibitemShut {NoStop}%
\bibitem [{\citenamefont {Eschenazi}, \citenamefont {Solari},\ and\
  \citenamefont {Gilmore}(1989)}]{eschenazi1989basins}%
  \BibitemOpen
  \bibfield  {author} {\bibinfo {author} {\bibfnamefont {E.}~\bibnamefont
  {Eschenazi}}, \bibinfo {author} {\bibfnamefont {H.~G.}\ \bibnamefont
  {Solari}},\ and\ \bibinfo {author} {\bibfnamefont {R.}~\bibnamefont
  {Gilmore}},\ }\bibfield  {title} {\enquote {\bibinfo {title} {Basins of
  attraction in driven dynamical systems},}\ }\href
  {https://doi.org/10.1103/PhysRevA.39.2609} {\bibfield  {journal} {\bibinfo
  {journal} {Physical Review A}\ }\textbf {\bibinfo {volume} {39}},\ \bibinfo
  {pages} {2609--2627} (\bibinfo {year} {1989})}\BibitemShut {NoStop}%
\bibitem [{\citenamefont {Soliman}\ and\ \citenamefont
  {Thompson}(1992)}]{soliman1992global}%
  \BibitemOpen
  \bibfield  {author} {\bibinfo {author} {\bibfnamefont {M.~S.}\ \bibnamefont
  {Soliman}}\ and\ \bibinfo {author} {\bibfnamefont {J.~M.~T.}\ \bibnamefont
  {Thompson}},\ }\bibfield  {title} {\enquote {\bibinfo {title} {Global
  dynamics underlying sharp basin erosion in nonlinear driven oscillators},}\
  }\href {https://doi.org/10.1103/PhysRevA.45.3425} {\bibfield  {journal}
  {\bibinfo  {journal} {Physical Review A}\ }\textbf {\bibinfo {volume} {45}},\
  \bibinfo {pages} {3425--3431} (\bibinfo {year} {1992})}\BibitemShut {NoStop}%
\bibitem [{\citenamefont {Grebogi}, \citenamefont {Ott},\ and\ \citenamefont
  {Yorke}(1986)}]{grebogi1986metamorphoses}%
  \BibitemOpen
  \bibfield  {author} {\bibinfo {author} {\bibfnamefont {C.}~\bibnamefont
  {Grebogi}}, \bibinfo {author} {\bibfnamefont {E.}~\bibnamefont {Ott}},\ and\
  \bibinfo {author} {\bibfnamefont {J.~A.}\ \bibnamefont {Yorke}},\ }\bibfield
  {title} {\enquote {\bibinfo {title} {Metamorphoses of basin boundaries in
  nonlinear dynamical systems},}\ }\href
  {https://doi.org/10.1103/PhysRevLett.56.1011} {\bibfield  {journal} {\bibinfo
   {journal} {Physical Review Letters}\ }\textbf {\bibinfo {volume} {56}},\
  \bibinfo {pages} {1011--1014} (\bibinfo {year} {1986})}\BibitemShut {NoStop}%
\bibitem [{\citenamefont {Wagemakers}, \citenamefont {Daza},\ and\
  \citenamefont {Sanju\'an}(2023)}]{wagemakers2023using}%
  \BibitemOpen
  \bibfield  {author} {\bibinfo {author} {\bibfnamefont {A.}~\bibnamefont
  {Wagemakers}}, \bibinfo {author} {\bibfnamefont {A.}~\bibnamefont {Daza}},\
  and\ \bibinfo {author} {\bibfnamefont {M.~A.~F.}\ \bibnamefont {Sanju\'an}},\
  }\bibfield  {title} {\enquote {\bibinfo {title} {Using the basin entropy to
  explore bifurcations},}\ }\href {https://doi.org/10.1016/j.chaos.2023.113963}
  {\bibfield  {journal} {\bibinfo  {journal} {Chaos, Solitons \& Fractals}\
  }\textbf {\bibinfo {volume} {175}},\ \bibinfo {pages} {113963} (\bibinfo
  {year} {2023})}\BibitemShut {NoStop}%
\bibitem [{\citenamefont {Datseris}(2018)}]{Datseris2018}%
  \BibitemOpen
  \bibfield  {author} {\bibinfo {author} {\bibfnamefont {G.}~\bibnamefont
  {Datseris}},\ }\bibfield  {title} {\enquote {\bibinfo {title}
  {Dynamicalsystems.jl: A {J}ulia software library for chaos and nonlinear
  dynamics},}\ }\href {https://doi.org/10.21105/joss.00598} {\bibfield
  {journal} {\bibinfo  {journal} {Journal of Open Source Software}\ }\textbf
  {\bibinfo {volume} {3}},\ \bibinfo {pages} {598} (\bibinfo {year}
  {2018})}\BibitemShut {NoStop}%
\bibitem [{\citenamefont {Schultz}\ \emph {et~al.}(2017)\citenamefont
  {Schultz}, \citenamefont {Menck}, \citenamefont {Heitzig},\ and\
  \citenamefont {Kurths}}]{schultz2017potentials}%
  \BibitemOpen
  \bibfield  {author} {\bibinfo {author} {\bibfnamefont {P.}~\bibnamefont
  {Schultz}}, \bibinfo {author} {\bibfnamefont {P.~J.}\ \bibnamefont {Menck}},
  \bibinfo {author} {\bibfnamefont {J.}~\bibnamefont {Heitzig}},\ and\ \bibinfo
  {author} {\bibfnamefont {J.}~\bibnamefont {Kurths}},\ }\bibfield  {title}
  {\enquote {\bibinfo {title} {Potentials and limits to basin stability
  estimation},}\ }\href {https://doi.org/10.1088/1367-2630/aa5a7b} {\bibfield
  {journal} {\bibinfo  {journal} {New Journal of Physics}\ }\textbf {\bibinfo
  {volume} {19}},\ \bibinfo {pages} {023005} (\bibinfo {year}
  {2017})}\BibitemShut {NoStop}%
\end{thebibliography}%

\end{document}